\documentclass[aps,reprint,nofootinbib,amsmath,amssymb]{revtex4-2}

\usepackage{amsmath,amssymb,amsfonts}
\usepackage{braket}
\usepackage{graphicx}
\usepackage{subfig}
\usepackage{float}
\usepackage{flafter}
\usepackage{booktabs}
\usepackage{siunitx}
\usepackage{listings}
\usepackage{algorithm}
\usepackage{algpseudocode}
\usepackage{xcolor}
\usepackage{hyperref}
\usepackage{xurl}
\usepackage{bm}
\usepackage{ragged2e}
\makeatletter
\long\def\@makecaption#1#2{%
  \par
  \vskip\abovecaptionskip
  \begingroup
   \footnotesize\rmfamily
   \justifying
   \setlength{\parindent}{0pt}%
   \let\footnote\@footnotemark@gobble
   \@make@capt@title{#1}{#2}\par
  \endgroup
  \vskip\belowcaptionskip
}
\makeatother
\hypersetup{
    colorlinks=true,
    linkcolor=blue,
    citecolor=blue,
    urlcolor=blue,
    pdfborder={0 0 0}
}

\graphicspath{{figures/}}

\begin{document}

\title{Molecular Dynamics-Derived Coloured Noise Mediates Anderson Localisation and Environment-Assisted Transport of Tryptophan Excitons in Tubulin}

\author{Chen Xin}
\email{imxinchen@outlook.com}


\date{July 15, 2026}

\begin{abstract}
The tryptophan residues in tubulin $\alpha\beta$-dimers form an ordered aromatic network that has been proposed to support quantum exciton transport even under physiological environmental noise. Existing studies of this system mostly assume white-noise dephasing, but the statistical properties of the protein--solvent bath coupled to tryptophan sites remain uncharacterised under physiological conditions. Here we characterise this fluctuation bath via all-atom molecular dynamics simulations of a solvated tubulin dimer at 310~K, combining high-frequency and long-time trajectories with 10~fs and 10~ps sampling intervals. The resulting autocorrelation of the site-energy fluctuations is tri-exponential, with three well-separated decay modes: sub-100-fs and picosecond fluctuations driven by water dynamics, and a nanosecond mode originating from protein conformational rearrangements. All three modes fall deep within the non-Markovian regime. We further demonstrate that the slow protein mode introduces strong quasi-static disorder, which results in Anderson localisation, while the two fast water modes frequently tune chromophore pairs through resonance, enabling environment-assisted quantum transport (ENAQT). On the full eight-site network, the coloured-noise bath confines excitons predominantly to strongly coupled proximal tryptophan pairs, in marked contrast to the more uniform delocalisation predicted by the standard white-noise Haken--Strobl model. Our workflow generalises to other pigment--protein systems with solvent-exposed chromophores.
\end{abstract}

\maketitle

\section{Introduction}
\label{sec:introduction}

The discovery of long-lived (660~fs) quantum coherence in the Fenna--Matthews--Olson (FMO) complex at cryogenic temperature (77~K)~\cite{engel_evidence_2007} ignited broad interest in quantum biology. Long assumed to wash out rapidly in warm and wet biological environments, quantum effects have been repeatedly observed under physiological conditions across photosynthetic systems: FMO coherence was subsequently observed at physiological temperature for at least 300~fs~\cite{panitchayangkoon_long-lived_2010}; long-lasting excitation oscillations have been detected in cryptophyte algae light-harvesting proteins~\cite{collini_coherently_2010}; ultrafast energy transfer pathways have been mapped in plant light-harvesting complex II (LHCII) at room temperature~\cite{wells_pathways_2014}; and exciton--vibrational coherence has been observed in cyanobacterial allophycocyanin~\cite{zhu_quantum_2024} for 501~fs, where the coherence is closely linked to molecular vibrations. These findings confirm that quantum effects can modulate biological exciton dynamics under native conditions.

Additionally, environment-assisted quantum transport (ENAQT) theory~\cite{mohseni_environment-assisted_2008} has shown that thermal fluctuations can actively assist transport when their amplitude and correlation time fall in the appropriate regime, refining the early intuition that the protein environment acts purely as a decoherence source. Ref.~\cite{chen_excitation_2011} established that the dependence is non-monotonic, with transfer maximised at an optimal intermediate noise correlation time, and Ref.~\cite{blach_environment-assisted_2025} recently confirmed this prediction experimentally in perovskite nanocrystal superlattices, where exciton diffusion peaks at a sample-dependent turnover temperature. The accuracy of environmental modelling is therefore a critical determinant of reliable theoretical predictions for excitonic systems.

Beyond photosynthetic antennae, the network of tryptophan residues in $\alpha\beta$-tubulin dimers constitutes another naturally ordered chromophore system for investigating biological excitonic transport. Microtubules, ubiquitous eukaryotic cytoskeletal filaments assembled from $\alpha\beta$-tubulin dimers, contain eight tryptophan residues per dimer. With the lowest-energy ${}^1L_a$ excited state among natural amino acids and a large transition dipole moment, tryptophan serves as a natural exciton carrier with physical mechanisms directly analogous to those in photosynthetic systems. Early debates about quantum effects in microtubules focused on charge polarisation and conformational superposition states, with substantial disagreement over decoherence timescales~\cite{tegmark_importance_2000, hagan_quantum_2002}; the tryptophan excitonic degree of freedom, by contrast, provides a physically grounded and experimentally testable framework.

The first quantitative eight-site Hamiltonian for this system was established in Ref.~\cite{craddock_feasibility_2014}, where exciton dynamics was treated under the Haken--Strobl white-noise approximation, a common simplification in the field.
Whether this simplification is justified depends strongly on the system: exact simulations show that neglecting the structured nature of the vibrational environment can bias electronic parameter estimates~\cite{caycedo-soler_exact_2022}, yet studies on the FMO complex find that Markovian descriptions may still adequately capture energy transfer dynamics due to the masking effect of the phonon background~\cite{mujica-martinez_quantification_2013}.
The degree of non-Markovianity and the applicability of white-noise models therefore cannot be assumed a priori, and all-atom molecular dynamics (MD) has emerged as a powerful tool to characterise native environmental fluctuations inaccessible to static structural analysis~\cite{liguori_light-harvesting_2015}.

Subsequent theoretical and experimental work has advanced our understanding of tubulin tryptophan excitons. Excitonic models extended to microtubule lattices predict superradiant lowest-energy exciton states and supertransfer~\cite{celardo_existence_2019}, and a recent Lindblad-based analysis of multi-spiral microtubule assemblies identified subradiant retention channels and signatures of non-Markovian information backflow~\cite{gassab_quantum_2026}. On the experimental side, tryptophan autofluorescence lifetime measurements revealed excitation diffusion lengths far exceeding Förster resonance energy transfer (FRET) predictions, which can be reversibly suppressed by volatile anaesthetics~\cite{kalra_electronic_2023}; steady-state fluorescence quantum yield measurements further corroborated collective excitonic behaviour, showing elevated quantum yields in polymerised microtubules compared to isolated dimers~\cite{babcock_ultraviolet_2024}.

However, existing theoretical studies of tubulin tryptophan excitons share a fundamental limitation: environmental modelling relies entirely on phenomenological assumptions without an atomistic foundation. While recent work has explored non-Markovianity arising from the structural connectivity of the microtubule lattice~\cite{gassab_quantum_2026}, the intrinsic non-Markovianity driven by the multi-timescale thermal fluctuations of the protein--solvent environment remains uncharacterised. Dephasing rates are artificially tuned, and the complex spectral density spanning four orders of magnitude is entirely ignored. No study to date has characterised the statistical properties and physical origins of tryptophan site-energy fluctuations directly from all-atom dynamics. Furthermore, although experiments suggest that environmental perturbations (e.g., by anaesthetics) alter energy transport via dielectric screening~\cite{kalra_electronic_2023}, the atomistic mechanism underlying this screening via protein--water electrostatic interactions remains unexplored.

In this work, we address these gaps by constructing a microscopic coloured-noise model for tryptophan site-energy fluctuations in the solvated tubulin dimer, derived entirely from all-atom MD simulations at 310~K with dual time-resolution. Our main findings are as follows. First, site-energy fluctuations follow a tri-exponential autocorrelation function with three well-separated relaxation timescales, all deep in the non-Markovian regime, demonstrating a qualitative failure of the Markovian approximation. Second, the slowest, protein-driven mode imposes strong quasi-static disorder and Anderson localisation, while faster water-driven modes break localisation and enable ENAQT, with native MD parameters falling within the optimal transport window. Third, source decomposition reveals strong protein--water electrostatic anticorrelation that suppresses effective disorder by a factor of $\sim\sqrt{2}$ via dielectric screening, providing a microscopic origin for tubulin's high optical dielectric constant. Fourth, on the full eight-site network, the coloured-noise bath confines excitons to strongly coupled proximal pairs, in contrast to the uniform delocalisation predicted by the white-noise model, explaining the suppressed fluorescence yield of isolated dimers through disorder-induced localisation. This work moves tubulin exciton modelling beyond phenomenological dephasing toward atomistically derived multi-timescale baths, and the general workflow is applicable to other pigment--protein systems.

The remainder of this paper is organised as follows. Section~\ref{sec:methods} details the computational methods, Sections~\ref{sec:results} and~\ref{sec:discussion} present the results and their physical implications, and Section~\ref{sec:conclusion} summarises our conclusions.

\section{Methods}
\label{sec:methods}

\subsection{Molecular Dynamics Simulation Setup}
\label{sec:methods-md}

The MD simulation was performed with GROMACS 2026.2~\cite{gromacs_2026}. The starting structure was taken from the Protein Data Bank as entry 1JFF~\cite{lowe_refined_2001}, which contains the $\alpha$ and $\beta$ monomers and three ligands: GTP, Mg\textsuperscript{2+}, and GDP. The dimer carries GTP at the non-exchangeable (N) site on the $\alpha$ monomer and GDP at the exchangeable (E) site on the $\beta$ monomer, with Mg\textsuperscript{2+} coordinating the GTP phosphates. The dimer contains eight tryptophan (Trp) residues, referred to as Trp1--Trp8 throughout this work following the convention of Ref.~\cite{craddock_feasibility_2014}: Trp1--4 are $\alpha$W21, $\alpha$W346, $\alpha$W388, and $\alpha$W407 on chain~A, and Trp5--8 are $\beta$W21, $\beta$W101, $\beta$W344, and $\beta$W397 on chain~B. The missing residues in the PDB entry (39 in chain~A and 19 in chain~B) were repaired with PDBFixer, and protonation states of ionisable residues were predicted with PROPKA at pH 7.0 via PDB2PQR~\cite{jurrus_improvements_2018}. The protein was described with the AMBER99SB-ILDN~\cite{lindorfflarsen_improved_2010} force field. GTP and GDP used the polyphosphate parameters from Ref.~\cite{meagher_development_2003}, which extend the AMBER parm94/99 family with atom charges and bonded terms for phosphate groups.\footnote{Parameter files obtained from the Bryce Group AMBER Parameter Database: \url{http://personalpages.manchester.ac.uk/staff/Richard.Bryce/amber/cof/}} The solvated system contained 128{,}053 atoms in a $12.39 \times 7.83 \times 13.31$~nm\textsuperscript{3} periodic box, with 37{,}976 TIP3P waters and 0.15~M NaCl. After steepest-descent energy minimisation to $F_{\max} < 100$~kJ~mol\textsuperscript{-1}~nm\textsuperscript{-1}, the system was equilibrated in two stages: 100~ps NVT followed by 500~ps NPT, both at 310~K and 1~bar with position restraints on heavy atoms. A 50~ns unrestrained NPT production run was then carried out with a 2~fs timestep, LINCS bond constraints, and PME electrostatics using a 1.0~nm real-space cutoff and 0.12~nm grid spacing. Frames were saved every 10~ps. The first 10~ns showed residual equilibration drift and was excluded from analysis, leaving 4{,}001 frames from the 10--50~ns window. To resolve sub-picosecond solvent dynamics that the 10~ps cadence cannot capture, an additional 2~ns continuation trajectory was run from the $t = 50$~ns endpoint with identical MD configuration but frames saved every 10~fs, yielding 200{,}001 frames. The two trajectories thus provide complementary time resolutions (Table~\ref{tab:trajectories}): the long trajectory captures nanosecond protein conformational dynamics, while the short trajectory resolves sub-picosecond solvent librations. The three ligands (GTP, Mg\textsuperscript{2+}, and GDP) remained bound at their respective sites throughout the 50~ns trajectory. A movie of the full simulation is available as supplementary material.

\begin{figure}[t]
    \centering
    \includegraphics[width=0.7\columnwidth]{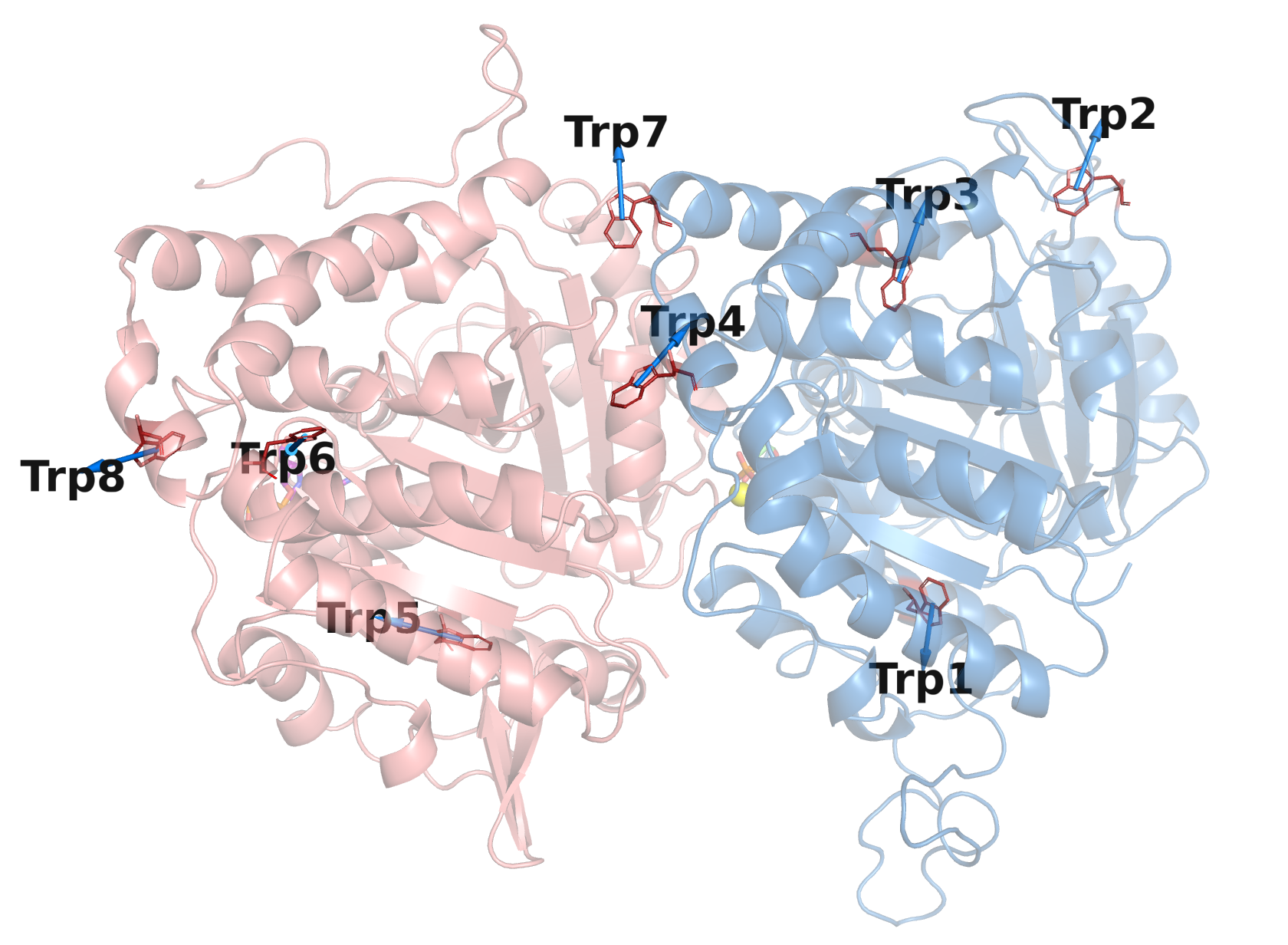}
    \caption{Dipole geometry of the eight tryptophans (Trp1--Trp8) in the tubulin $\alpha\beta$-dimer, extracted from the MD frame at $t = 50$~ns. Each blue arrow starts at the centroid of an indole ring and points along $\bm{\mu}_m$ (from the benzene ring toward the pyrrole ring, reflecting the direction of the S$_0 \to {}^1L_a$ difference dipole). The dipole magnitudes are assumed identical ($\mu = 5$~D) for all Trp residues, but the directions $\bm{n}_m(t)$ fluctuate with the trajectory. The Trp1--Trp8 labels follow Ref.~\cite{craddock_feasibility_2014}: Trp1--4 are $\alpha$W21, $\alpha$W346, $\alpha$W388, $\alpha$W407 on chain~A, and Trp5--8 are $\beta$W21, $\beta$W101, $\beta$W344, $\beta$W397 on chain~B.}
    \label{fig:md_setup}
\end{figure}

\begin{table}[t]
    \centering
    \setlength{\tabcolsep}{8pt}
    \caption{The two MD trajectories used in this work. They share identical MD configurations (2~fs integrator step, deep thermal equilibration) and differ only in output cadence $\Delta t$ for data efficiency, and they resolve complementary timescales of $\delta\epsilon_m(t)$.}
    \label{tab:trajectories}
    \begin{tabular}{lrrrr}
        \toprule
        trajectory & $\Delta t$ & frames & span & Nyquist \\
        \midrule
        long & 10~ps  & 4\,001   & 40~ns & 1.67~cm$^{-1}$ \\
        short & 0.01~ps  & 200\,001 & 2~ns & 1668~cm$^{-1}$ \\
        \bottomrule
    \end{tabular}
\end{table}

\subsection{Site-energy Fluctuation Modelling}
\label{sec:methods-site-energy}
The site-energy fluctuation model is based on the linear Stark effect and the excitation Hamiltonian from Ref.~\cite{craddock_feasibility_2014}.

The site energy of the $m$-th tryptophan in a tubulin dimer is the energy needed to excite the tryptophan from $S_0$ to ${}^1L_a$, the lowest fluorescent excited state of tryptophan under physiological conditions. Ref.~\cite{craddock_feasibility_2014} formulates the site energy $\epsilon_m$ by three terms:
\begin{equation}
    \epsilon_m = \epsilon_0 + \epsilon_{\mathrm{qm}} + \epsilon_{\mathrm{coul}},
\end{equation}
where $\epsilon_0 = 35000$~cm\textsuperscript{-1} is a constant baseline transition energy for spectral alignment. The quantum correction term $\epsilon_{\mathrm{qm}}$ accounts for the intrinsic vertical excitation energy of each isolated Trp residue, obtained from scaled time-dependent density-functional theory (TD-DFT) calculations. The electrostatic contribution $\epsilon_{\mathrm{coul}}$ describes the environmentally induced electrochromic shift, evaluated via charge-density coupling (CDC) between the differential excitation charge density of each Trp and the background protein electrostatic field. In Ref.~\cite{craddock_feasibility_2014}, the site energy is treated as static for subsequent exciton dynamics simulations performed via the Haken--Strobl pure-dephasing model.

To capture site-energy fluctuations induced by the thermal environment, we introduce time dependence to the site energies. We assume the quantum correction term $\epsilon_{\mathrm{qm}}$ is time-invariant, so the temporal variation of the site energy arises predominantly from the electrostatic term $\epsilon_{\mathrm{coul}}$. This approximation is well justified: small-amplitude side-chain rotations induce only minor variations in the intrinsic vertical excitation energy (on the order of tens of cm\textsuperscript{-1}), which are far smaller than the electrostatically induced site-energy fluctuations ($10^2-10^3$~cm\textsuperscript{-1}) and therefore contribute negligibly.

$\epsilon_{\mathrm{coul}}(t)$ is evaluated within the linear Stark approximation, which involves only the transition difference dipole moment $\bm{\mu}_m$ of the indole ring of Trp $m$ and the local electric field $\bm{E}_m(\bm{r}_m, t)$ at the indole ring center $\bm{r}_m$:
\begin{align}
    \epsilon_{\mathrm{coul}}(t) &= -\bm{\mu}_m \cdot \bm{E}_m(\bm{r}_m, t) \notag \\
        &= -\mu \Big(\bm{n}_m(t) \cdot \bm{E}_m(\bm{r}_m, t)\Big).
\end{align}
Notice that here $\bm{\mu}_m$ is decomposed further into its magnitude $\mu$ (assumed constant for all Trp residues) and its direction $\bm{n}_m(t)$, as the orientational fluctuation dominates over the magnitude variation. The site-energy fluctuation of the $m$-th Trp is then given by
\begin{align}
    \delta\epsilon_m(t) &= \epsilon_m(t) - \langle \epsilon_m(t) \rangle_T \notag \\
        &= -\mu \Big(\bm{n}_m \cdot \bm{E}_m - \langle \bm{n}_m \cdot \bm{E}_m \rangle_T\Big),
\end{align}
where $\langle \cdot \rangle_T$ denotes the time average over the MD trajectory.

The magnitude $\mu$ for tryptophan residues in protein interiors has been reported in the range of 4.82--8~D, and the electron density shifts from the pyrrole ring to the benzene ring upon excitation from $S_0$ to ${}^1L_a$~\cite{vivian_mechanisms_2001}. We adopt a representative value of $\mu = 5$~D for all tryptophan residues, and use unit vectors pointing from the benzene to the pyrrole ring to represent the direction of the difference dipole moment. Figure~\ref{fig:md_setup} shows the geometry of the extracted dipole directions in one MD frame.

To compute the site-energy fluctuation, the local electric field $\bm{E}_m$ at the indole ring center $\bm{r}_m$ is extracted from the MD trajectory and projected onto the dipole direction $\bm{n}_m(t)$.  The electric field $\bm{E}_m$ is computed by direct Coulomb summation over the explicit MD partial charges:
\begin{align}
    \bm{E}_m(\bm{r}_m, t) = \frac{1}{4\pi\varepsilon_0}\sum_{i=1}^{K} q_i\, e \, \frac{\bm{r}_m - \bm{r}_i^{(\mathrm{bg})}}{\lvert \bm{r}_m - \bm{r}_i^{(\mathrm{bg})} \rvert^3},
    \label{eq:efield}
\end{align}
where $\varepsilon_0$ is the vacuum permittivity, $e$ is the elementary charge, $q_i$ are the AMBER99SB-ILDN partial charges (in units of $e$) at positions $\bm{r}_i^{(\mathrm{bg})}$ of the background environment, $K$ is the total number of background atoms. The summation excludes all atoms belonging to the indole side chain of Trp $m$, while including the protein backbone, other tryptophan residues, solvent molecules, nucleotides, and ions. The background charges are partitioned into four disjoint groups: protein, water, nucleotide, and ions. Therefore, Eq.~\ref{eq:efield} can be evaluated separately for each source, allowing the fluctuation $\delta\epsilon_m(t)$ to be decomposed accordingly.

The dielectric treatment requires careful explanation. In the static CDC site-energy calculation of Ref.~\cite{craddock_feasibility_2014}, only protein partial charges enter the Coulomb sum, and the missing solvent screening is compensated by an effective optical dielectric constant $\varepsilon_{\mathrm{opt}} = 8.41$, the experimentally measured high-frequency dielectric of tubulin. In the MD-based approach, the trajectory explicitly resolves all sources of electrostatic response (solvent, ions, nucleotides, and protein), so the instantaneous fluctuation of $\bm{E}_m$ is captured intrinsically by the rearranged positions of these explicit partial charges, and no additional continuum dielectric is applied ($\varepsilon_r = 1$, i.e., vacuum permittivity).

The two treatments are physically complementary. We retain the static Hamiltonian from Ref.~\cite{craddock_feasibility_2014}, including mean site energies and inter-Trp couplings computed with $\varepsilon_{\mathrm{opt}} = 8.41$, as the baseline describing static energetic disorder. The MD-derived fluctuation $\delta\epsilon_m(t)$ is added on top to account for dynamic disorder induced by thermal motions. Inter-site excitonic couplings are treated as static throughout this work, as their relative fluctuations are substantially smaller than site-energy disorder and thus negligible.

\subsection{Exciton Dynamics}
\label{sec:methods-hamiltonian}

Under the single-excitation approximation, the exciton Hamiltonian
is expressed in the site basis $\{\ket{m}\}_{m=1}^{8}$ as
\begin{align}
H(t) &= H_0 + \sum_m \delta\epsilon_m(t) \ket{m}\bra{m} \notag \\
     &= \sum_m (\epsilon_m + \delta\epsilon_m(t)) \ket{m}\bra{m} 
        + \sum_{m \neq n} J_{mn} \ket{m}\bra{n}
\end{align}
where $\epsilon_m$ is the site energy of the $m$-th tryptophan,
$J_{mn}$ is the excitonic coupling between sites $m$ and $n$,
and $\delta\epsilon_m(t)$ denotes the MD-derived site-energy fluctuations
described in Sec.~\ref{sec:methods-site-energy}.
Both $\epsilon_m$ and $J_{mn}$ are taken from Ref.~\cite{craddock_feasibility_2014}.

The dynamics of each realisation are governed by the stochastic Schr\"odinger equation (SSE):
\begin{align}
i\hbar|\dot{\psi}(t)\rangle = H(t)|\psi(t)\rangle.
\end{align}
Because $H(t)$ is stochastic, physical observables are obtained by ensemble-averaging over many independent realisations.
The bath is fully characterised by its autocorrelation function (ACF).
Assuming the process is stationary (verified in Appendix~\ref{sec:appendix-stationarity}),
the ACF depends only on the lag $\tau = t - t'$:
\begin{align}
C_m(\tau) = \langle \delta\epsilon_m(t)\, \delta\epsilon_m(t + \tau) \rangle_T.
\end{align}
The ACF is well described by a tri-exponential decay with three
well-separated timescales $T_k$ ($k{=}1,2,3$), as determined in
Sec.~\ref{sec:results-timescales}.
All three modes lie deep in the non-Markovian regime
(Kubo number $\kappa_k = \sigma_k T_k / \hbar \gg 1$~\cite{kubo_stochastic_1963},
a dimensionless measure of bath memory; values reported in
Sec.~\ref{sec:results-exciton}), requiring a coloured-noise treatment
rather than a Markovian white-noise approximation.

\begin{figure}[t]
    \centering
    \includegraphics[width=\columnwidth]{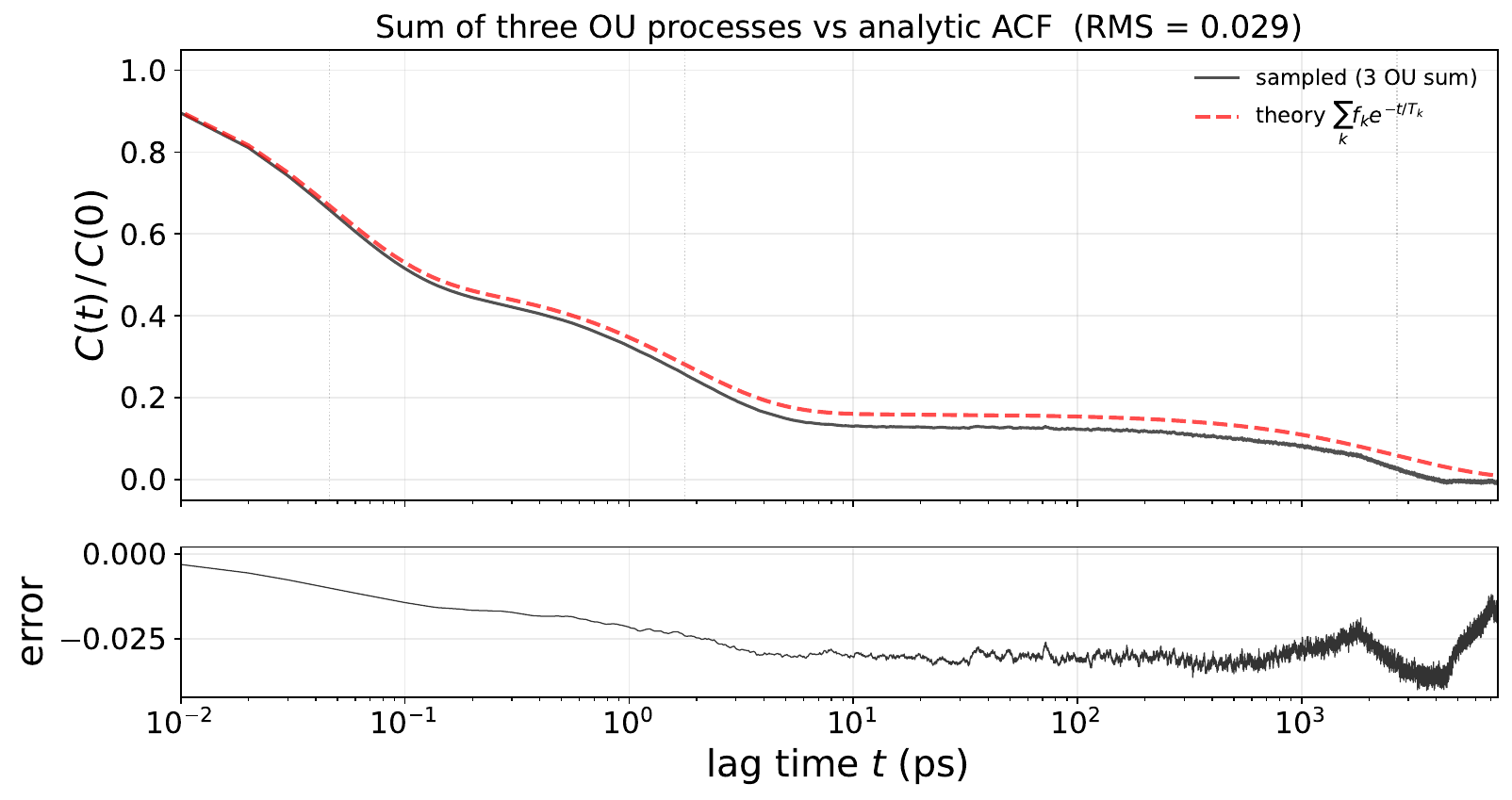}
    \caption{ACF of the OU sum (a single 30~ns trajectory, five-realisation average; dots) versus the analytic target $\sum_k f_k e^{-t/T_k}$ (solid). \textbf{Lower panel:} residual (sampled $-$ theory). RMS residual $=0.028$ over 10~fs--7.5~ns, confirming the sampler reproduces the fitted statistics.}
    \label{fig:ou_verify}
\end{figure}

We therefore model the bath as a superposition of three independent
Ornstein--Uhlenbeck (OU) processes, one per timescale:
\begin{align}
\delta\epsilon_m(t) &= \sum_{k=1}^{3} x_{m,k}(t)
\end{align}
where $x_{m,k}$ is an OU process with ACF
\begin{align}
\langle x_{m,k}(t)\, x_{m,k}(t + \tau) \rangle_T = \sigma_{m,k}^2 \mathrm{e}^{-|\tau| / T_k}
\end{align}
Each process can be sampled exactly using the Gillespie discretisation scheme~\cite{gillespie_exact_1996}:
\begin{align}
x_{m,k}(t + \mathrm{d} t) = &x_{m,k}(t) e^{-\mathrm{d} t / T_k} \notag \\
                        &+ \sigma_{m,k} \sqrt{1 - e^{-2\mathrm{d} t / T_k}} \, \xi(t),
\end{align}
where $\xi(t)$ is a random variable sampled from the standard normal distribution $\mathcal{N}(0,1)$. Fig.~\ref{fig:ou_verify} shows the ACF computed from the summation of the three sampled OU processes and the tri-exponential fit to the MD-derived ACF, confirming that the OU sampler reproduces the fitted statistics. The total fluctuation variance on each site is then $\sigma_m^2 = \sum_k \sigma_{m,k}^2$.

The SSE is solved by Monte Carlo (MC) trajectory sampling.
For each of $N_{\mathrm{MC}} = 500$ independent bath realisations,
the stochastic Schrödinger equation $i\hbar|\dot\psi\rangle = H(t)|\psi\rangle$
is integrated numerically with an adaptive-step ODE solver,
with the OU noise entering $H(t)$ as time-dependent site-diagonal
coefficients sampled at $\mathrm{d} t = 2$~fs.
The ensemble-averaged density matrix
$\bar\rho(t) = N_{\mathrm{MC}}^{-1}\sum_i |\psi_i(t)\rangle\langle\psi_i(t)|$
yields the physical observables.
The resulting coloured-noise model (CNM) is compared against the
Haken--Strobl model (HSM)~\cite{rips_stochastic_1993, chen_effect_2010},
a $\delta$-correlated white-noise reference.
Computational parameters and results are given in
Sec.~\ref{sec:results-exciton}.
Numerical propagation uses QuTiP v5.3.0~\cite{qutip5}.

\section{Results}
\label{sec:results}

We now characterise the site-energy fluctuations $\delta\epsilon_m(t)$ from two MD trajectories of complementary time resolution (Sec.~\ref{sec:methods-md}--\ref{sec:methods-site-energy}, Table~\ref{tab:trajectories}), with excitonic couplings and mean site energies from Ref.~\cite{craddock_feasibility_2014}. Prior dynamical studies on this Hamiltonian have treated the environment within the Haken--Strobl white-noise framework; the MD-derived bath presented below lies instead deep in the non-Markovian regime, and Sec.~\ref{sec:results-exciton} quantifies the resulting differences in exciton transport.

This section is structured as follows. First, we quantify the overall amplitude and statistical properties of the site-energy fluctuations (Sec.~\ref{sec:results-magnitude}). Next, we dissect the multiple timescales of the underlying bath modes (Sec.~\ref{sec:results-timescales}), decompose fluctuations by physical origin, and evaluate dielectric screening contributions from each environment component (Sec.~\ref{sec:results-source}). We conclude by presenting exciton dynamics simulations under the coloured-noise bath, first on the strongly-coupled Trp4--Trp7 two-site system and then on the full eight-site network (Sec.~\ref{sec:results-exciton}).

\subsection{Magnitude and statistics of the fluctuations}
\label{sec:results-magnitude}

\begin{table*}[t]
    \centering
    \caption{Per-site fluctuation characterisation. $\sigma$: standard deviation of the site-energy fluctuations from the long trajectory, taken as the total amplitude. $\sigma_{\mathrm{short}}$: standard deviation from the short trajectory, which undersamples the nanosecond slow mode and is therefore systematically smaller (mean $\sigma^2/\sigma_{\mathrm{short}}^2\approx 1.2$). $\sigma/J$: ratio of $\sigma$ to the strongest coupling $|J_{47}|=59$~cm$^{-1}$. $f_k,T_k$: weights and timescales of the corrected tri-exponential fit to the autocorrelation (Sec.~\ref{sec:results-timescales}); all times in ps.}
    \label{tab:bath_summary}
    \small
    \setlength{\tabcolsep}{6pt}
    \begin{tabular}{lrrr rr rr rr}
        \toprule
        site & $\sigma$ (cm$^{-1}$) & $\sigma_{\mathrm{short}}$ (cm$^{-1}$) & $\sigma/J$ & $f_1$ & $T_1$ (ps) & $f_2$ & $T_2$ (ps) & $f_3$ & $T_3$ (ps) \\
        \midrule
        Trp1 &  844 &  782 & 14.3 & 0.56 & 0.036 & 0.36 & 0.51 & 0.07 & 1926 \\
        Trp2 &  972 &  755 & 16.5 & 0.46 & 0.027 & 0.40 & 0.64 & 0.14 &  776 \\
        Trp3 &  806 &  636 & 13.7 & 0.62 & 0.068 & 0.14 & 1.77 & 0.24 & 6820 \\
        Trp4 &  999 &  861 & 16.9 & 0.57 & 0.041 & 0.37 & 1.00 & 0.05 & 8524 \\
        Trp5 &  721 &  726 & 12.2 & 0.47 & 0.050 & 0.13 & 0.41 & 0.39 &  957 \\
        Trp6 &  846 &  947 & 14.3 & 0.53 & 0.062 & 0.31 & 8.58 & 0.16 &  492 \\
        Trp7 &  675 &  627 & 11.4 & 0.52 & 0.034 & 0.38 & 0.58 & 0.09 &  452 \\
        Trp8 & 1453 & 1512 & 24.6 & 0.36 & 0.047 & 0.52 & 0.65 & 0.12 & 1358 \\
        \midrule
        \textbf{mean} & \textbf{914} & \textbf{856} & \textbf{15.5} & 0.51 & 0.046 & 0.33 & 1.77 & 0.16 & 2663 \\
        \bottomrule
    \end{tabular}
\end{table*}

The site-energy fluctuations $\delta\epsilon_m(t)$ are zero-mean by construction, approximately Gaussian, and stationary (Appendix~\ref{sec:appendix-stationarity}). The standard deviation from the long trajectory $\sigma_{\mathrm{long}}$ is systematically larger than that from the short trajectory $\sigma_{\mathrm{short}}$ (Table~\ref{tab:bath_summary}), because the 2~ns short trajectory resolves the sub-picosecond bath but undersamples the nanosecond slow mode, whereas the 40~ns long trajectory captures both. Therefore, we take $\sigma_{\mathrm{long}}$ as the total fluctuation amplitude, denoted as $\sigma$ for simplicity. Also notice that the ratio $\sigma/J$ ranges from 11 (Trp7) to 25 (Trp8), placing every site deep in the \textbf{strong-disorder} regime ($\sigma/J\gg 1$), for which Anderson localisation of the exciton is expected.

\subsection{Three well-separated relaxation timescales}
\label{sec:results-timescales}

\begin{figure}[t]
    \centering
    \includegraphics[width=\columnwidth]{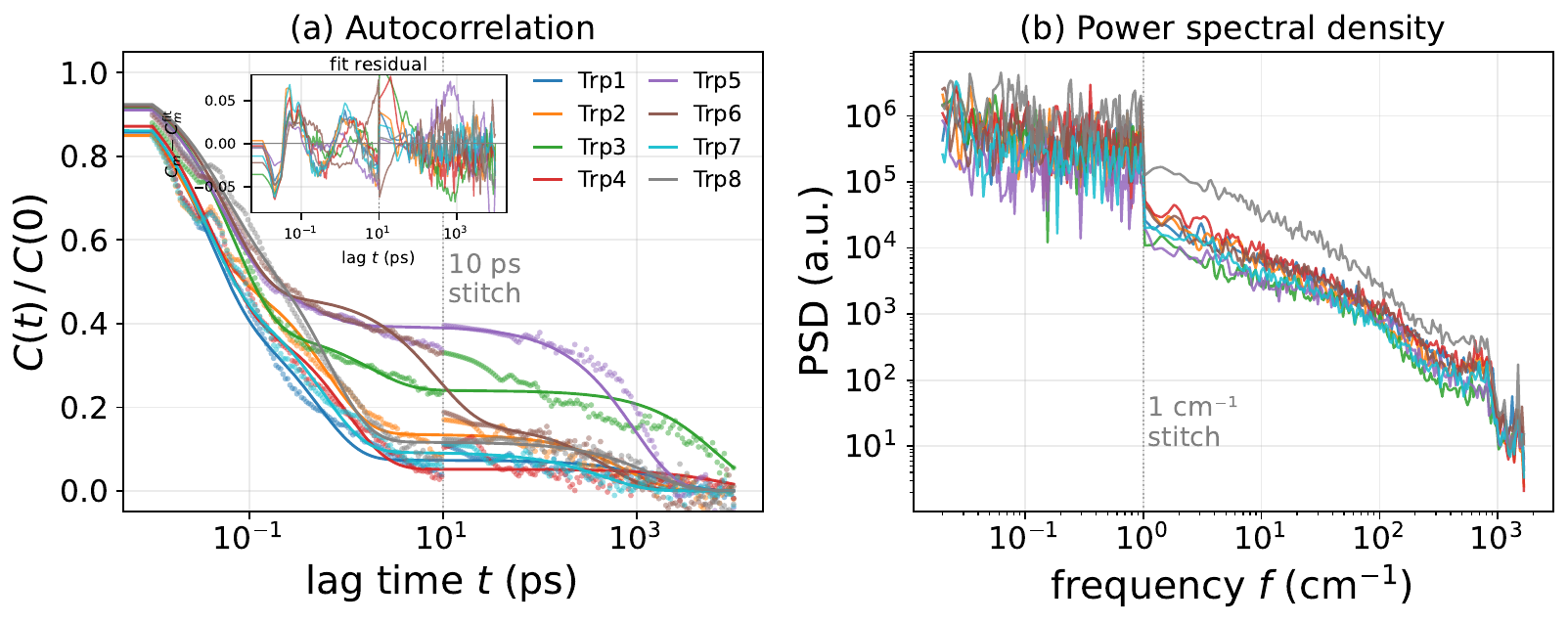}
    \caption{Stitched autocorrelation and power spectrum of the site-energy fluctuations. \textbf{(a)} Self-normalised autocorrelations $C_m(t)$ (dots) with corrected tri-exponential fits (solid lines; two-step refit with $T_3$ anchored to the long trajectory, Sec.~\ref{sec:results-timescales}); per-site $R^2 \ge 0.972$ (mean $0.987$). \emph{Inset}: residual $C_m - C_m^{\mathrm{fit}}$ vs lag, showing the spike at the 10~ps stitch (grey dotted), the known signature of slow-mode undersampling in the short trajectory. \textbf{(b)} Stitched power spectral densities (arbitrary units). The stitch is set at $f=1$~cm$^{-1}$, below the long trajectory's Nyquist frequency ($1.67$~cm$^{-1}$, determined by its $10$~ps sampling). The visible jump at the stitch is due to aliasing: power from frequencies above the Nyquist frequency folds back into the $[0,\,1.67]$~cm$^{-1}$ band, artificially inflating the long-trajectory PSD.}
    \label{fig:acf_psd}
\end{figure}

Autocorrelation functions (ACFs) for each tryptophan site $C_m(t)=\langle\delta\epsilon_m(0)\delta\epsilon_m(t)\rangle/\sigma_m^2$ are computed independently on both trajectories and \emph{stitched} at lag time $t=10$~ps (the long trajectory's first lag): the short ACF for $t<10$~ps, the long ACF for $t\ge 10$~ps. The power spectral densities (PSDs) are stitched analogously at $f=1$~cm$^{-1}$. Because each ACF is self-normalised, the per-trajectory variance mismatch is absorbed into the normalisation; absolute variance re-enters downstream computations via $\sigma_m$.

The stitched ACFs are decisively described by a \textbf{tri-exponential} decay,
\begin{equation}
    C_m(t)=\sum_{k=1}^{3} f_k\,e^{-t/T_k},\qquad \textstyle\sum_{k=1}^{3} f_k=1,
    \label{eq:triexp}
\end{equation}
rather than a bi-exponential one. We select between the two with the Akaike information criterion~\cite{akaike_new_1974} $\mathrm{AIC}=n\ln(\mathrm{RSS}/n)+2k$ ($n$: number of data points; RSS: residual sum of squares; $k$: number of free parameters), whose $2k$ term penalises free parameters so that a model cannot win merely by overfitting. On the stitched ACF the tri-exponential lowers the AIC by $\Delta\mathrm{AIC}=-331$ relative to the bi-exponential, giving decisive evidence, since $|\Delta\mathrm{AIC}|>10$ is conventionally regarded as strong support for the lower-AIC model. This comparison fixes only the model structure (three components rather than two); the parameter values themselves are extracted below with explicit treatment of the 10~ps stitch artefact, which would otherwise bias the slow timescale.

The stitched ACF is not smooth at the 10~ps crossover: the short trajectory underestimates the nanosecond-mode variance ($\sigma^2_{\mathrm{long}}/\sigma^2_{\mathrm{short}}\approx 1.2$, Table~\ref{tab:bath_summary}), so its self-normalised ACF sits below the long-trajectory ACF at the stitch. Directly fitting a tri-exponential across this discontinuity yields a high global $R^2=0.988$ but systematically underestimates $T_3$, because the optimiser bends the slow decay faster to absorb the upward jump at 10~ps; the resulting $T_3\approx 1.1$~ns is less than half the unbiased value of $2.66$~ns. The fit residual is localised at the stitch, where it reaches $0.04$--$0.07$ per site (2--4$\times$ the typical residual of ${\sim}0.02$ elsewhere; Fig.~\ref{fig:acf_psd}(a), inset). We therefore adopt a two-step procedure: (i) read $T_3$ from the long trajectory alone, which resolves the nanosecond mode, and (ii) refit the two fast components on the stitched ACF with $T_3$ anchored to this value. The global $R^2$ remains $0.987$ ($\Delta R^2=-0.001$), confirming that the correction removes the $T_3$ bias without degrading the overall fit. The residual at 10~ps is thus the known signature of slow-mode undersampling in the short trajectory, not a failure of the tri-exponential model itself. Fig.~\ref{fig:acf_psd}(a) shows the corrected tri-exponential fits for all eight Trps, and Table~\ref{tab:bath_summary} reports the fitted parameters.

The physics picture underlying the three timescales $T_1$, $T_2$, $T_3$ is well understood from the literature (e.g. Ref.~\cite{laage_perspective_2017}). 
\begin{itemize}
    \item $T_1\sim 46$~fs ($f_1=0.51$): sub-100-fs librational motions of water molecules strongly coupled to the protein surface.
    \item $T_2\sim 1.77$~ps ($f_2=0.33$): water reorientation and hydrogen-bond reformation.
    \item $T_3\sim 2.66$~ns ($f_3=0.16$): nanosecond protein conformational dynamics and tumbling; slow and site-dependent, reflecting heterogeneous local environments.
\end{itemize}
Here $f_1$, $f_2$, $f_3$ indicate each mode's fractional contribution to the total fluctuation variance $\sigma_m^2$. We see that the two fast modes dominate the variance (84\%), while the slow mode contributes only 16\%. 
These timescales will be further validated by the source decomposition analysis below.

\subsection{Source decomposition and dielectric screening}
\label{sec:results-source}

To decompose the site-energy fluctuation by environmental source, we first verify that the difference dipole $\bm{\mu}_m$ is effectively static on the exciton timescale. 
Its autocorrelation $C_{\bm{\mu}}(\tau)=\langle \bm{\mu}_m(t) \cdot \bm{\mu}_m(t+\tau) \rangle_t$ remains $0.984$ after a 2~ps lag on the short trajectory (site-averaged; per-site ranges $0.98$--$0.99$).
Consistently, freezing $\bm{\mu}_m$ to its time-average $\bar{\bm{\mu}}_m$ and recomputing $\delta\epsilon_m$ leaves the site-averaged standard deviation unchanged ($\sigma_{\mathrm{fixed}}/\sigma=1.00$, specifically $0.998$ on the long trajectory and $1.003$ on the short). The fluctuation is therefore carried by the field $\bm{E}_m(t)$ alone, and we hold $\bm{\mu}_m=\bar{\bm{\mu}}_m$ fixed in the following analysis.

With $\bm{\mu}_m$ fixed, $\delta\epsilon_m(t)=-\bar{\bm{\mu}}_m\cdot\bm{E}_m(t)$ is linear in the field. The Coulomb field decomposes exactly by source group, $\bm{E}_m=\sum_g\bm{E}_{g,m}$ ($g\in\{$protein, water, nucleotide, ions$\}$), so the site energy inherits the same decomposition, $\delta\epsilon_m=\sum_g\delta\epsilon_{g,m}$. Fig.~\ref{fig:source_decomp}(a) shows the decomposed $\sigma_m$, and Fig.~\ref{fig:source_decomp}(b) shows the decomposed PSDs. Two findings stand out:

\begin{figure}[t]
    \centering
    \includegraphics[width=\columnwidth]{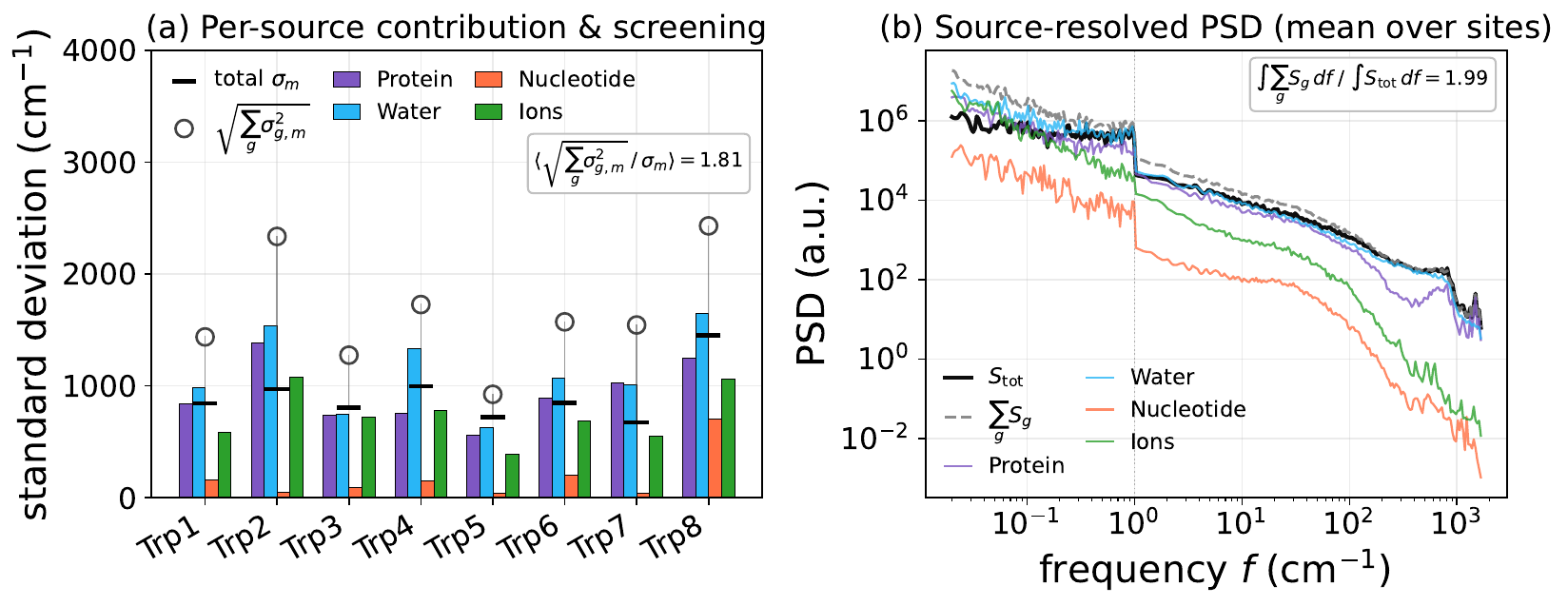}
    \caption{Source decomposition and dielectric screening. \textbf{(a)} Per-source standard deviation $\sigma_{g,m}$ (bars) with the measured total $\sigma_m$ (black marker) and the uncorrelated expectation $\sqrt{\sum_g\sigma_{g,m}^2}$ (open circles); the gap between the two markers is the cross-source cancellation. \textbf{(b)} Source-resolved PSD, averaged over the eight sites, with the measured total (black) and the direct source sum (grey dashed). Water dominates every frequency band, protein is second, and nucleotide is negligible. The direct sum overestimates the total by a factor of two due to cross-source cancellation.}
    \label{fig:source_decomp}
\end{figure}

\paragraph{Water dominates every frequency band.}
Water is the largest fluctuation source across all bands, contributing 52--65\% of the spectral power in each (54\% in the slow band). Protein is consistently second (27--39\%), while nucleotide is negligible everywhere ($<1.3\%$). Ions contribute $\sim$18\% only in the slow band and vanish at sub-ps frequencies.

\paragraph{Dielectric screening.}
The component spectra sum to \emph{twice} the measured total ($\int\sum_g S_g\,df\,/\,\int S_{\mathrm{tot}}\,df=2.0$). Equivalently, the screening ratio $R_{\mathrm{screen}}=1-\sigma_m^2/\sum_g\sigma_{g,m}^2$ averages $0.66$, meaning cross-source terms cancel two-thirds of the naive variance sum. The dominant cancellation is protein--water, whose fluctuations are anti-correlated (Pearson $r=-0.39$) because polarised water partially cancels the protein field at the indole rings. Any noise model that treats sources as independent therefore overestimates $\sigma$ by $\sim\sqrt{2}$.

Fitting the tri-exponential to each source's own ACF independently confirms the timescale assignment of Sec.~\ref{sec:results-timescales}. Water recovers the two fast modes ($T_1\approx 54$~fs, $T_2\approx 1.1$~ps), while protein recovers the nanosecond mode ($T_3\approx 2.5$~ns). The fast modes are thus water-driven and the slowest mode protein-driven; the temporal and source decompositions independently converge on the same physical picture.

\subsection{Exciton dynamics under coloured noise}
\label{sec:results-exciton}

We now apply the CNM and HSM of Sec.~\ref{sec:methods-hamiltonian} to the tryptophan network, using the Hamiltonian from Ref.~\cite{craddock_feasibility_2014} reproduced in Table~\ref{tab:hamiltonian}.

\begin{table}[t]
    \centering
    \footnotesize
    \setlength{\tabcolsep}{4pt}
    \caption{Exciton Hamiltonian (cm$^{-1}$) of the eight Trp sites in the tubulin dimer, copied from Ref.~\cite{craddock_feasibility_2014}. Diagonal entries are site-energy offsets relative to a base of $35888$~cm$^{-1}$; off-diagonal entries are excitonic couplings $J_{mn}$.}
    \label{tab:hamiltonian}
    \begin{tabular}{c|rrrrrrrr}
        \toprule
         & 1 & 2 & 3 & 4 & 5 & 6 & 7 & 8 \\
        \midrule
        Trp1 & $1$ & $0$ & $-13$ & $0$ & $-2$ & $-1$ & $5$ & $-1$ \\
        Trp2 & $0$ & $388$ & $-41$ & $4$ & $1$ & $1$ & $-4$ & $1$ \\
        Trp3 & $-13$ & $-41$ & $342$ & $2$ & $0$ & $1$ & $-6$ & $1$ \\
        Trp4 & $0$ & $4$ & $2$ & $207$ & $-4$ & $6$ & $-59$ & $-1$ \\
        Trp5 & $-2$ & $1$ & $0$ & $-4$ & $57$ & $21$ & $2$ & $11$ \\
        Trp6 & $-1$ & $1$ & $1$ & $6$ & $21$ & $102$ & $5$ & $-51$ \\
        Trp7 & $5$ & $-4$ & $-6$ & $-59$ & $2$ & $5$ & $248$ & $3$ \\
        Trp8 & $-1$ & $1$ & $1$ & $-1$ & $11$ & $-51$ & $3$ & $0$ \\
        \bottomrule
    \end{tabular}
\end{table}

\begin{figure}[t]
    \centering
    \includegraphics[width=\columnwidth]{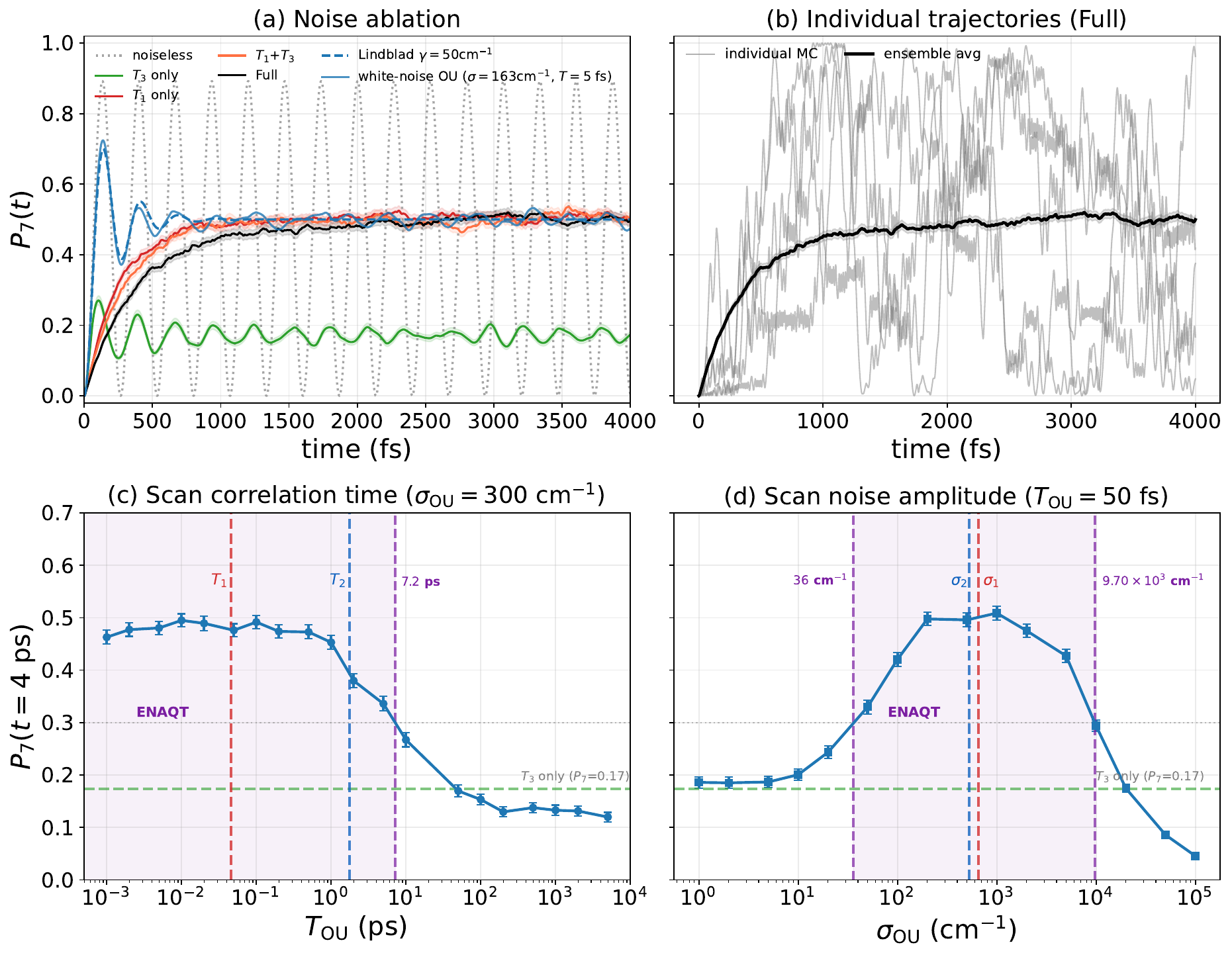}
    \caption{Inhomogeneous dephasing and ENAQT on the Trp4--Trp7 two-site system ($J = -59~\mathrm{cm}^{-1}$).
    The excitation is initialised on Trp4 in all panels. Each site carries its own bath of three additive Ornstein--Uhlenbeck processes fitted to the per-site MD fluctuations (Table~\ref{tab:bath_summary}).
    Dynamics are computed by MC sampling of the stochastic Schr\"odinger equation with $N_{\mathrm{MC}} = 500$ realisations.
    \textbf{(a) Noise ablation:} population $P_7(t)$ on Trp7 under various noise configurations. The noiseless case (grey dotted) gives coherent Rabi oscillations. The slowest mode $T_3$ alone partially localises the excitation ($P_7(4~\mathrm{ps}) = 0.17$, green). Adding the fast mode $T_1$ breaks the localisation and enables transport ($T_1$ alone: $P_7 = 0.50$; $T_1{+}T_3$: $P_7 = 0.50$; full model: $P_7 = 0.50$, black), demonstrating ENAQT. For solver validation, the HSM solution at $\gamma=50~\mathrm{cm}^{-1}$ (blue dashed) is compared with a white-noise MC trajectory ($\sigma_{\mathrm{OU}}=163~\mathrm{cm}^{-1}$, $T_{\mathrm{OU}}=5$~fs, tuned so that $2\sigma^2 T=\gamma$; blue solid). 
    \textbf{(b) Individual trajectories:} five MC realisations under the full bath (grey) together with the $N_{\mathrm{MC}} = 500$ ensemble average (black). Individual trajectories oscillate coherently, whereas the ensemble average dephases smoothly. This is the signature of inhomogeneous dephasing, where coherence is lost by phase dispersion across realisations rather than by within-trajectory dissipation.
    \textbf{(c) ENAQT scan over correlation time:} $P_7(4~\mathrm{ps})$ vs.\ $T_{\mathrm{OU}}$, with $T_3$ always present at full MD strength and an additional OU process of fixed amplitude $\sigma_{\mathrm{OU}} = 300~\mathrm{cm}^{-1}$ scanned over its correlation time. Defining the ENAQT regime as $P_7 > 0.3$ (dotted line), transport is assisted for all $T_{\mathrm{OU}} \lesssim 7$~ps and is optimal near $\hbar/J \approx 90$~fs. The MD $T_1$ mode ($\sim$35--41~fs, red markers) sits in the optimal regime; $T_2$ ($\sim$0.4--1.0~ps at most sites, blue) lies well within.
    \textbf{(d) ENAQT scan over noise amplitude:} as in (c) but with $T_{\mathrm{OU}} = 50$~fs fixed and $\sigma_{\mathrm{OU}}$ scanned. The ENAQT window ($P_7 > 0.3$) spans $\sigma_{\mathrm{OU}} \approx 36$--$9700~\mathrm{cm}^{-1}$, with a broad peak near $\sigma_{\mathrm{OU}} \sim 200$--$1000~\mathrm{cm}^{-1}$ ($P_7 \approx 0.51$). Weaker noise cannot overcome the $T_3$ static disorder ($\sigma_3\sim 200$~cm$^{-1}$) to reach resonance; excessively strong noise ($\sigma_{\mathrm{OU}}\gg J$, Kubo number $\kappa\gg 1$) creates large quasi-static detunings that keep the system far from resonance.}
    \label{fig:exciton}
\end{figure}

First, we focus on an isolated pair Trp4--Trp7, which has the strongest coupling $|J|=59$~cm$^{-1}$ and a site-energy offset $\Delta\epsilon=41$~cm$^{-1}$. Each site carries its three-component OU bath (Table~\ref{tab:bath_summary}). For each of $N_{\mathrm{MC}}=500$ realisations the Hamiltonian
\begin{equation}
    H(t) = \begin{bmatrix}
        \epsilon_4 + \delta\epsilon_4(t) & J \\[4pt]
        J & \epsilon_7 + \delta\epsilon_7(t)
    \end{bmatrix}
    \label{eq:hsys}
\end{equation}
is propagated unitarily with time step $\Delta t=2$~fs for an observation time $T_{\mathrm{obs}}=4$~ps in total. The observables are then ensemble-averaged. This computation is exact for classical Gaussian noise. The two sites are treated as independent, justified by the negligible spatial correlations (Appendix~\ref{sec:appendix-corr}). Since the triple OU processes give us plenty of room to tweak, a noise-ablation simulation is conducted, as shown in Fig.~\ref{fig:exciton}. 

As can be seen from Fig.~\ref{fig:exciton}(a), with only the slow component $T_3$ ($\sigma_3 = 227~\mathrm{cm}^{-1}$ on Trp4, $205~\mathrm{cm}^{-1}$ on Trp7), the exciton remains partially localised on Trp4 ($P_4\approx 0.83$ at 4~ps).
Since $T_3$ far exceeds the observation window, the system is under strong static disorder, resulting in Anderson localisation~\cite{anderson_absence_1958}. However, after adding the much faster $T_1$ and/or $T_2$ modes, the localisation is broken and the population on Trp7 rises to $\sim 0.50$ at 4~ps. This is a clear demonstration of ENAQT~\cite{PhysRevLett.122.050501,Moix_2013,rebentrost_environment-assisted_2009}, where moderate dynamic noise counteracts Anderson localisation to enhance transport. The phenomenon has recently been directly observed in experiments on perovskite nanocrystal superlattices, where exciton transport is maximised at intermediate dephasing strength~\cite{blach_environment-assisted_2025}. In our system, the fast noise drives the detuning across resonance repeatedly, opening transfer windows for the exciton to hop from Trp4 to Trp7. In addition, the individual MC realisations (Fig.~\ref{fig:exciton}(b), grey) show coherent oscillations which are averaged out by the ensemble averaging. This is a signature of inhomogeneous dephasing. 

To further investigate the ENAQT effect, we first apply the $T_3$ mode noise to Trp4 and Trp7 to place the two-site system in the strong static-disorder regime, then scan over the correlation time $T_{\mathrm{OU}}$ or amplitude $\sigma_{\mathrm{OU}}$ of an additional OU process (fixing one and varying the other). ENAQT is considered effective when $P_7$ exceeds 0.3 at 4~ps. Fig.~\ref{fig:exciton}(c) shows that, at fixed amplitude $\sigma_{\mathrm{OU}}=300~\mathrm{cm}^{-1}$, correlation times below $\sim 7$~ps enable ENAQT, with maximum transfer near $T_{\mathrm{OU}}\lesssim 1$~ps. Both the $T_1$ and $T_2$ modes characterised in Sec.~\ref{sec:results-timescales} fall within this regime. Fig.~\ref{fig:exciton}(d) shows the complementary scan over amplitude at fixed $T_{\mathrm{OU}}=50$~fs. The ENAQT regime spans $\sigma_{\mathrm{OU}}\approx 36$--$9700~\mathrm{cm}^{-1}$, with a broad peak near $\sigma_{\mathrm{OU}}\sim 200$--$1000~\mathrm{cm}^{-1}$. Weaker noise ($\sigma_{\mathrm{OU}}\lesssim 36$~cm$^{-1}$) cannot overcome the $T_3$ static disorder ($\sigma_3\sim 200$~cm$^{-1}$ per site) to bring the detuning across resonance, while excessively stronger noise ($\sigma_{\mathrm{OU}}\gg J$) creates large quasi-static detunings (Kubo number $\kappa=\sigma T/\hbar\approx 90$ at the upper boundary; $\kappa\gg 1$ marks the non-Markovian regime where the bath acts quasi-statically and the white-noise approximation fails), so the system rarely visits resonance.

An OU process reaches the Haken--Strobl white-noise limit when its correlation time tends to zero with the product $\sigma^2 T$ held fixed, so the same Monte Carlo code, run in this limit, must reproduce the analytic HSM solution. We exploit this to validate the solver: a single symmetric OU process per site with $T=5$~fs and $\sigma\approx 163$~cm$^{-1}$, tuned so that $2\sigma^2 T=\gamma=50$~cm$^{-1}$ ($1/\gamma=106$~fs; the factor of 2 accounts for both sites dephasing the coherence symmetrically), is propagated as a white-noise MC and overlaid in Fig.~\ref{fig:exciton}(a) alongside the analytic HSM transient for comparison. The MC reproduces the HSM transient ($P_7(4~\mathrm{ps})=0.47\pm 0.01$ versus the HSM value of $0.50$), confirming the solver. The matching pair $(\sigma, T) = (163~\mathrm{cm}^{-1}, 5~\mathrm{fs})$ is not a physical bath but simply the OU representation of the phenomenological rate $\gamma=50$~cm$^{-1}$. For comparison, the CNM's fast mode alone ($\sigma_1\approx 650$~cm$^{-1}$, $T_1\approx 46$~fs) would correspond to a Markovian dephasing rate $\sim\!10^2\gamma$ if it were white noise, yet the coloured bath produces comparable transfer because its spectral power is distributed away from the system frequency by the large Kubo number $\kappa_1\approx 6$.

Although the two-site system clearly illustrates the ENAQT mechanism, it cannot distinguish the coloured-noise model from white-noise dephasing. The endpoint carries no discriminating information, because any two-level system with pure dephasing approaches an equal-population limit of $1/2$ at long times regardless of the underlying mechanism or transient shape. We therefore turn to the full eight-site network, which provides a more discriminating test: its steady-state population distribution directly reveals the difference between delocalising white-noise dephasing and confinement by quasi-static disorder.

\begin{figure}[t]
    \centering
    \includegraphics[width=\columnwidth]{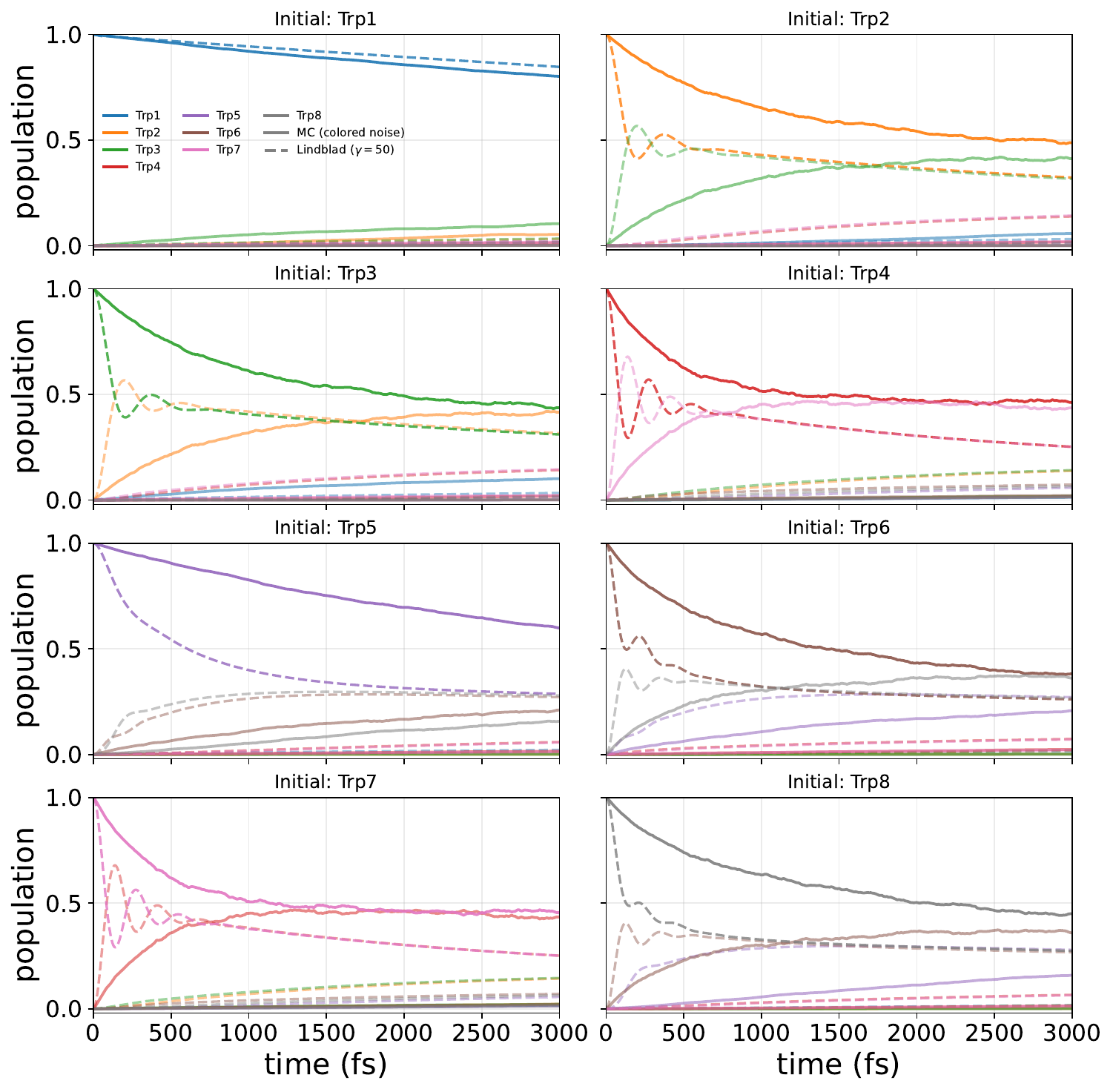}
    \caption{Exciton dynamics on the full eight-site Trp network~\cite{craddock_feasibility_2014}, for initial excitation on each of the eight sites (3~ps evolution, $N_{\mathrm{MC}}=500$). In every panel, solid lines show the CNM and dashed lines the HSM at $\gamma=50$~cm$^{-1}$; colours distinguish the eight Trp sites. The CNM consistently retains more population on the starting site and its strongest-coupled partner (e.g.\ Trp4$\leftrightarrow$Trp7), whereas the HSM spreads population across the full network.}
    \label{fig:exciton_8site}
\end{figure}

Fig.~\ref{fig:exciton_8site} shows the population dynamics under the HSM (dashed) and the CNM (solid) for initial excitation on each of the eight Trp sites. In every panel the CNM retains more population on the starting site and its strongest-coupled partner than the HSM. Starting from Trp4, for instance, the CNM gives $P_4(3~\mathrm{ps})=0.46$, $P_7(3~\mathrm{ps})=0.44$, and only $P_{\mathrm{leak}}=0.10$ escapes the pair. The HSM, by contrast, spreads population across the full network ($P_{\mathrm{leak}}=0.50$), approaching the long-time uniform limit of $1/8$ per site.

This qualitative difference is expected because the MD-derived bath violates the assumptions of the HSM on every count: (i)~the bath is slow, with even the fastest correlation time $T_1\approx 46$~fs comparable to the coherent timescale $\hbar/J_{\max}\approx 90$~fs (and $T_2$, $T_3$ exceeding it by 1--4 orders of magnitude); and (ii)~the coupling is strong, with $\sigma/J\approx 11$--$25$ at every site. Crucially, all three modes are deep in the non-Markovian regime ($\kappa_1\sim 6$, $\kappa_2\sim 10^2$, $\kappa_3\sim 10^5$, all $\gg 1$). The HSM dephasing rate $\gamma=50$~cm$^{-1}$ is therefore not an \emph{ab initio} prediction of the bath but a phenomenological fit, and exciton transport on the tryptophan network proceeds via ENAQT under a bath that lies outside the HSM's validity.

Collectively, this section establishes a multi-scale picture of tryptophan site-energy disorder in a tubulin dimer system, drawn from all-atom molecular dynamics at 310~K in explicit solvent. The site-energy fluctuations follow a tri-exponential autocorrelation spanning solvent librational ($T_1\approx 46$~fs), water reorientation ($T_2\approx 1.77$~ps), and protein conformational ($T_3\approx 2.66$~ns) timescales, with water dominating the spectral power and substantial protein--water dielectric screening ($R_{\mathrm{screen}}\approx 0.66$). The observed ENAQT emerges from the competition between these bath components: the quasi-static $T_3$ disorder localises excitons through inhomogeneous dephasing and Anderson localisation, while the faster $T_1$ and $T_2$ modes dynamically modulate the site detuning and repeatedly drive the system through resonance, opening transient transfer windows. Transport peaks at an intermediate fast-noise strength that matches the MD-derived $T_1$ and $T_2$ timescales, and falls off when the noise is too weak to overcome the static localisation or so strong that it reintroduces quasi-static detunings. With all three components deep in the non-Markovian regime ($\kappa_k\gg 1$), the full eight-site network confirms the picture: the CNM confines excitons to strongly coupled site pairs, whereas the Markovian HSM delocalises them across the entire network.

\section{Discussion}
\label{sec:discussion}

\paragraph{The bath is non-Markovian.}
This work fundamentally revisits the theoretical treatment of tryptophan exciton dynamics in tubulin, which has relied on the Haken--Strobl white-noise approximation, e.g., \cite{craddock_feasibility_2014, gassab_quantum_2026}. Our dual-time-resolution MD of the solvated tubulin dimer at 310~K delivers a microscopically derived coloured-noise bath, demonstrating that every relaxation mode resides deep in the non-Markovian regime (Kubo number $\kappa_k \gg 1$), where white-noise and perturbative approximations fail qualitatively~\cite{chen_excitation_2011}. On the full network, the HSM tends to overestimate delocalisation, spreading population broadly, while the coloured bath confines 90\% of the excitation within spatially proximal pairs.

\paragraph{The noise can be a friend.}
The MD-derived noise model exhibits a tri-exponential autocorrelation, revealing three physically distinct modes: (i) sub-100-fs water librations, (ii) picosecond water reorientation and hydrogen-bond rearrangements, and (iii) nanosecond protein conformational motions. At 310~K, the slowest mode imposes strong static disorder on the tryptophan network, resulting in Anderson localisation~\cite{anderson_absence_1958}, while the two fast modes dynamically drive chromophore pairs through resonance, enabling ENAQT~\cite{PhysRevLett.122.050501,Moix_2013,rebentrost_environment-assisted_2009} and facilitating exciton hopping between proximal tryptophan pairs. This illustrates that environmental fluctuations are not inherently detrimental to exciton transport but can, under appropriate conditions of amplitude and timescale, actively facilitate energy transfer. The competition between static disorder and ENAQT observed here is consistent with prior theoretical and experimental results. For example, Ref.~\cite{chen_excitation_2011} showed that an optimal noise correlation time maximises exciton transport efficiency in FMO systems. Ref.~\cite{blach_environment-assisted_2025} showed experimentally that the mean squared displacement of excitons in perovskite nanocrystal superlattices is non-monotonous in temperature, peaking at a sample-dependent turnover temperature where static disorder and dephasing are balanced. These results collectively indicate that coherence, static disorder, and thermal fluctuations interact intricately to govern exciton transport, opening new possibilities for controlling exciton dynamics in biological and synthetic systems.

\paragraph{Dielectric screening.}
Dielectric screening acquires a clear microscopic interpretation within our all-atom framework. By decomposing the electric field at each tryptophan into contributions from protein, water, nucleotide, and ions, we reveal strong anticorrelation between the protein and water components (Pearson $r = -0.39$), which cancels two-thirds of the naive variance sum ($R_{\mathrm{screen}} \approx 0.66$) and reduces the total disorder amplitude by a factor of $\sim\sqrt{2}$ relative to independent-source estimates. Any noise model that treats these sources as independent therefore overestimates the effective disorder. Crucially, this dynamic anticorrelation arises from the same protein--water polarisation response that gives tubulin its unusually high optical dielectric constant $\varepsilon_{\mathrm{opt}} = 8.41$~\cite{mershin_tubulin_2004}, a value adopted in prior tubulin exciton calculations~\cite{craddock_feasibility_2014} through the local-field framework~\cite{juzeliunas_quantum_1994}. Our atomistic decomposition thus reveals that the static local-field enhancement of inter-chromophore couplings and the dynamic suppression of site-energy fluctuations are two manifestations of the same polarisation mechanism, unifying the equilibrium and fluctuational descriptions of the dielectric environment.

\paragraph{Exciton localisation and fluorescence.}
In the noise-free Hamiltonian (Table~\ref{tab:hamiltonian}) of the tryptophan network in tubulin, the intrinsic site-energy spread (388~cm$^{-1}$) already limits the maximum participation ratio of the eight tryptophan sites to 2.07, and a disorder-strength scan shows that the MD-derived quasi-static component ($\sigma_3 \approx 200$~cm$^{-1}$) further reduces it to $\sim$1.4 (see Appendix~\ref{sec:appendix-pr}). Such strong localisation precludes collective radiative enhancement in the isolated dimer. This is consistent with the measurements of Ref.~\cite{babcock_ultraviolet_2024}, who reported fluorescence quantum yields of 12.4\% for free tryptophan, 10.6\% for the tubulin dimer, and 17.6\% for assembled microtubules, and attributed the depressed dimer yield to non-radiative protein-environment channels. Our analysis identifies disorder-induced localisation as a complementary, independent mechanism. Upon microtubule assembly, collective inter-dimer coupling broadens the exciton band and may overcome this quasi-static disorder, plausibly explaining the elevated microtubule yield.

\paragraph{Limitations and outlook.}
Several simplifications frame the scope of this work. First, the fluctuation autocorrelation is obtained by stitching a short high-resolution trajectory with a long low-resolution one; the discontinuity at the stitch introduces some fitting uncertainty in the tri-exponential parameters. Second, the excitonic couplings $J_{mn}$, taken from Ref.~\cite{craddock_feasibility_2014}, are based on the point-dipole approximation with a uniform empirical optical dielectric constant. Third, the difference dipole magnitude is fixed at $\mu = 5$~D, whereas the indole ${}^1L_a$ value ranges from 4.82 to 8~D~\cite{vivian_mechanisms_2001}. Nevertheless, this choice does not undermine our core physical conclusions: because $\delta\epsilon_m\propto\mu$, the disorder amplitudes $\sigma_m$ and Kubo numbers $\kappa_k$ all scale linearly with $\mu$, so the dimensionless ratios governing the physics rescale by the single factor $\mu/(5~\mathrm{D})$. Even at the lower bound ($\mu = 4.82$~D), every site remains deep in the strong-disorder regime ($\sigma/J \geq 11 \gg 1$, Table~\ref{tab:bath_summary}), and increasing $\mu$ only strengthens the Anderson localisation. Looking forward, the full pipeline generalises to other pigment--protein systems, including photosynthetic light-harvesting complexes (LHC) and FMO complexes, and future refinements such as polarisable force fields, extension to the microtubule lattice, and dynamic coupling fluctuations will further improve the atomistic accuracy of the model.

\section{Conclusions}
\label{sec:conclusion}

This work establishes a microscopic coloured-noise model for tryptophan excitons in tubulin by deriving all bath statistics directly from dual-time-resolution all-atom molecular dynamics at 310~K. Our atomistic simulation resolves site-energy fluctuations across four orders of magnitude in time, from sub-100-fs solvent libration to nanosecond protein conformational dynamics, and decomposes them into three well-separated relaxation modes that all reside deep in the non-Markovian regime ($\kappa_k \gg 1$). This replaces the phenomenological dephasing rate of the Haken--Strobl white-noise model with a microscopic, multi-timescale bath derived entirely from protein and solvent dynamics.

The central physical message is that environmental noise is not inherently detrimental to exciton transport but can actively facilitate it. The slowest mode ($T_3$) imposes strong quasi-static disorder and Anderson localisation, while the two fast modes ($T_1$, $T_2$) repeatedly drive chromophore pairs through resonance, enabling ENAQT. On the full eight-site network, this competition confines excitons within strongly coupled tryptophan pairs, in marked contrast to the uniform delocalisation predicted by the Haken--Strobl model. Protein--water dielectric screening further suppresses the effective disorder by a factor of $\sim\sqrt{2}$, providing a microscopic origin for the optical dielectric constant adopted in prior static calculations.

These findings move chromophore-disorder modelling beyond phenomenological dephasing toward fully atomistic, multi-timescale environmental baths. The workflow generalises to any pigment--protein complex with water-interfaced chromophores, offering a route to quantify non-Markovian disorder in biological excitonic systems.

\begin{acknowledgments}
All research design, molecular dynamics simulations, data analysis, and manuscript preparation were completed independently by the author. This work builds on the theoretical training obtained during the author’s M.Sc. studies in Physics at the National University of Singapore. CUDA-accelerated GROMACS simulations were performed on rented GPU computational resources from AI Galaxy (\url{https://ai-galaxy.cn}). The GLM 5.1 large language model was used to accelerate the preparation of GROMACS input parameter configurations and preliminary data analysis scripts; all generated outputs were manually inspected, verified, and revised by the author to guarantee physical consistency and numerical accuracy. All analysis codes are publicly available on GitHub at \url{https://github.com/Varato/tubulin-bath-fluctuation}; the raw molecular dynamics trajectories ($\sim$100~GB) are available from the author upon reasonable request. The author expresses sincere gratitude to family members for their persistent support and encouragement throughout this study.
\end{acknowledgments}

\bibliographystyle{apsrev4-2}
\bibliography{ref}

\clearpage

\appendix
\setcounter{table}{0}
\setcounter{figure}{0}
\renewcommand{\thetable}{A\arabic{table}}
\renewcommand{\thefigure}{A\arabic{figure}}

\onecolumngrid
\begin{center}
    \large\bfseries Appendix
\end{center}
\vspace{3ex}
\twocolumngrid

\section{Spatial correlations of site-energy fluctuations}
\label{sec:appendix-corr}

\begin{figure}[t]
    \centering
    \includegraphics[width=0.6\columnwidth]{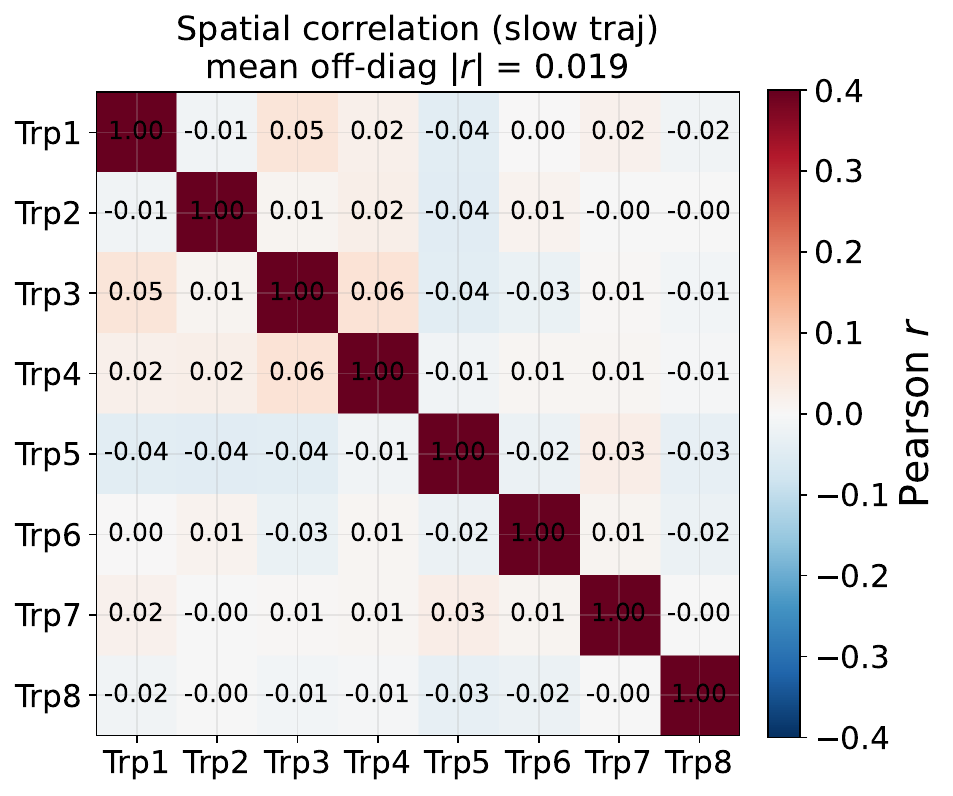}
    \caption{$8\times 8$ Pearson correlation matrix of the site-energy fluctuations (long trajectory). Mean off-diagonal $|r|=0.019$. Black lines mark the chain boundary (Trp1--4: $\alpha$-tubulin; Trp5--8: $\beta$-tubulin).}
    \label{fig:spatial_corr}
\end{figure}

Figure~\ref{fig:spatial_corr} shows the $8\times 8$ Pearson correlation matrix of the site-energy fluctuations over the long trajectory. All off-diagonal entries are small (mean $|r|=0.019$); the sites are statistically independent, justifying the diagonal covariance used in the exciton-dynamics treatment of Sec.~\ref{sec:results-exciton}.

\section{Fluctuation traces}
\label{sec:appendix-traces}

\begin{figure}[t]
    \centering
    \includegraphics[width=\columnwidth]{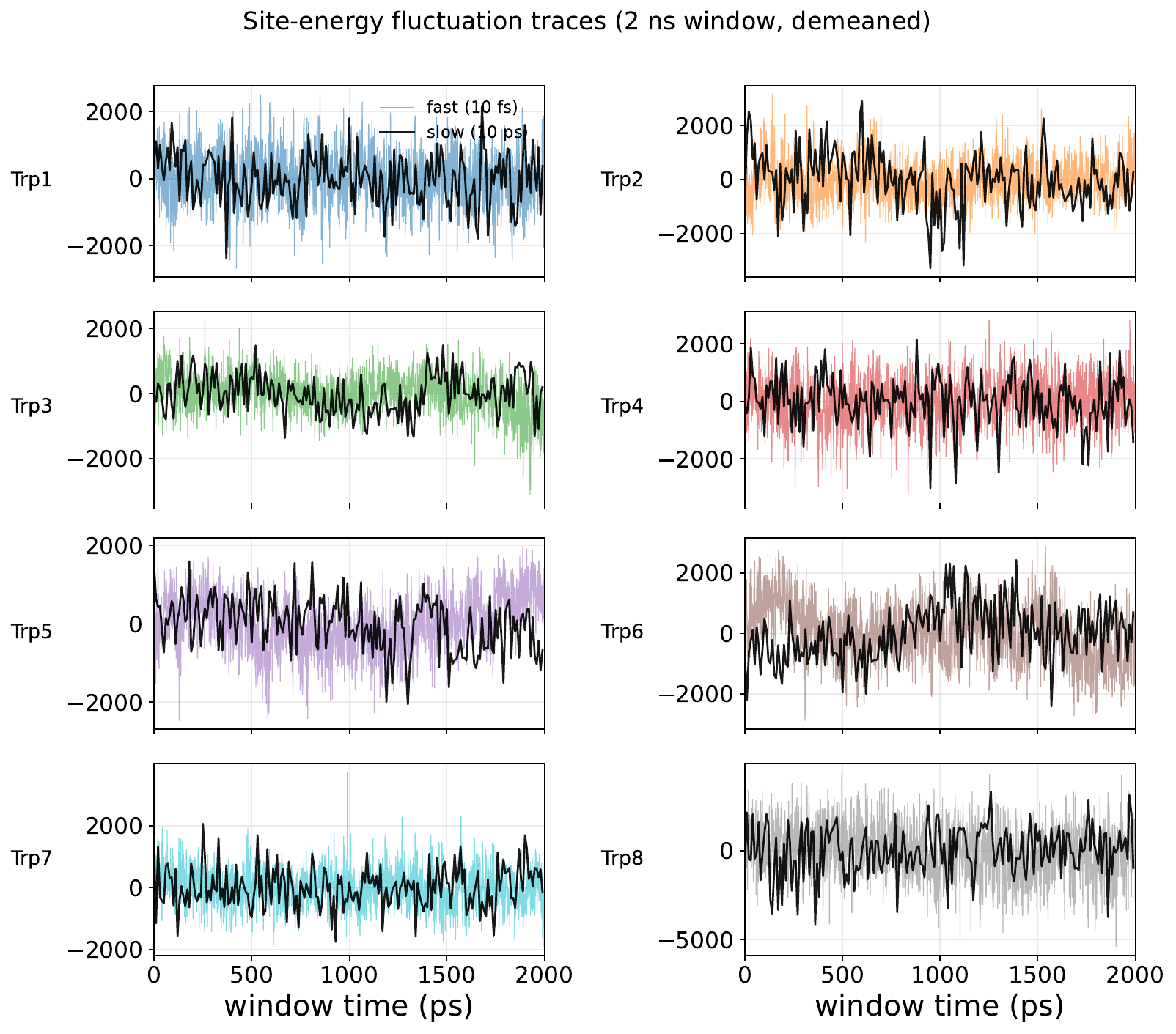}
    \caption{Two-nanosecond windows of $\delta\epsilon_m(t)$ for each Trp. The short trajectory (colour, 10~fs cadence) is overlaid with the last 2~ns of the long trajectory (black, 10~ps cadence), both demeaned. The coarse long trace resolves the nanosecond envelope; the short trace resolves the sub-picosecond jitter invisible at 10~ps sampling.}
    \label{fig:traces}
\end{figure}

Figure~\ref{fig:traces} shows representative windows of the raw $\delta\epsilon_m(t)$ that underlie the statistics of Sec.~\ref{sec:results-magnitude}. The traces are visually consistent with stationary fluctuation about a fixed mean; the amplitude contrast between sites mirrors the $\sigma_m$ spread of Table~\ref{tab:bath_summary}.

\section{Distribution and stationarity of the site-energy fluctuations}
\label{sec:appendix-stationarity}

\paragraph{Zero mean.}
The quantity analysed throughout this work is the fluctuation about the time-averaged site energy,
$\delta\epsilon_m(t) = -\bm{\mu}_m\!\cdot\!\bm{E}_m(t) - \langle -\bm{\mu}_m\!\cdot\!\bm{E}_m\rangle_T$.
The raw Stark shift carries a large, site-dependent static mean (the average electrostatic site energy, of order $10^2$--$10^3$~cm\textsuperscript{-1}); subtracting it makes the fluctuation exactly zero-mean, $\langle\delta\epsilon_m\rangle_T = 0$, by construction. Its standard deviation therefore coincides with its root-mean-square, and we report $\sigma_m = \mathrm{std}(\delta\epsilon_m)$ throughout.

\paragraph{Gaussianity.}
Table~\ref{tab:gaussianity} reports the skewness $\gamma$ and excess kurtosis $\kappa$ of $\delta\epsilon_m$ on both trajectories. Both are small on each ($|\gamma|\le 0.30$ long, $\le 0.32$ short; $|\kappa|\le 1.1$ long, $\le 0.7$ short), so the marginal distribution is close to Gaussian. This justifies treating the autocorrelation as a complete second-order descriptor of the bath and, later, synthesising the bath by Gaussian Monte-Carlo sampling.

\begin{table}[t]
    \centering
    \caption{Skewness $\gamma$ and excess kurtosis $\kappa$ of the zero-mean fluctuation $\delta\epsilon_m$ (a Gaussian has $\gamma=\kappa=0$), for both trajectories.}
    \label{tab:gaussianity}
    \small
    \begin{tabular}{lcccc}
        \toprule
         & \multicolumn{2}{c}{skewness $\gamma$} & \multicolumn{2}{c}{excess kurtosis $\kappa$} \\
        \cmidrule(lr){2-3}\cmidrule(lr){4-5}
        site & long & short & long & short \\
        \midrule
        Trp1 & $-0.15$ & $-0.05$ & $+0.24$ & $+0.06$ \\
        Trp2 & $-0.30$ & $+0.32$ & $+0.78$ & $+0.48$ \\
        Trp3 & $-0.29$ & $-0.32$ & $+0.54$ & $+0.68$ \\
        Trp4 & $-0.13$ & $-0.26$ & $-0.04$ & $+0.12$ \\
        Trp5 & $-0.03$ & $-0.10$ & $+0.29$ & $-0.13$ \\
        Trp6 & $-0.02$ & $+0.03$ & $-0.07$ & $-0.11$ \\
        Trp7 & $-0.08$ & $+0.11$ & $+1.07$ & $+0.62$ \\
        Trp8 & $-0.11$ & $-0.20$ & $+0.05$ & $-0.07$ \\
        \bottomrule
    \end{tabular}
\end{table}

\paragraph{Stationarity.}
Treating the autocorrelation as a function of lag only, $C_m(\tau)$, and pooling the long trajectory into a single ensemble require that the analysis window (10--50~ns) sample a stationary distribution. We verify this by split-half comparison: the 40~ns window is divided into two 20~ns segments (10--30~ns and 30--50~ns) and the per-segment standard deviation $\sigma_m$ compared site by site. The two halves agree to within 15\% for all eight Trps, with no systematic drift; the largest off-diagonal spatial correlation is likewise stable (0.090 and 0.076 in the two halves versus 0.056 over the full window, the half-window values slightly larger as expected from smaller-sample fluctuations). The first 10~ns of the 50~ns production run, which showed residual equilibration drift, was excluded from all analyses (Sec.~\ref{sec:methods-md}). The site-energy fluctuations are stationary over the analysis window.

\section{Structural descriptors}
\label{sec:appendix-structure}

\begin{table}[t]
    \centering
    \caption{Structural descriptors of the eight Trp sites. SASA: time-averaged whole-residue solvent-accessible surface area (Lee--Richards). $\theta_{\mathrm{loc}}$: Procrustes-corrected local angular deviation of the difference dipole --- the orientational analogue of RMSF, computed after removing the per-frame global rotation $R(t)$ (Wahba's problem on the 8-vector dipole set) and averaging $\theta_m(t)=\arccos(\hat{\bm{n}}_m^{\mathrm{loc}}(t)\cdot\langle\hat{\bm{n}}_m^{\mathrm{loc}}\rangle)$. $d_{\mathrm{nucl}}$: distance to the nearest GTP/GDP. Hydration: mean count of water molecules within 3.5~\AA\ of the indole ring.}
    \label{tab:structure}
    \small
    \begin{tabular}{lcccc}
        \toprule
        site & SASA (nm\textsuperscript{2}) & $\theta_{\mathrm{loc}}$ (deg) & nearest nucl.\ (\AA) & hydration \\
        \midrule
        Trp1 & 0.194 & 12.9 & 14.2 (GTP) & 1.71 \\
        Trp2 & 0.703 & 15.0 & 29.2 (GTP) & 2.19 \\
        Trp3 & 0.046 & 12.7 & 14.5 (GTP) & 0.13 \\
        Trp4 & 0.488 &  8.3 & 12.2 (GTP) & 2.36 \\
        Trp5 & 0.029 &  8.6 & 13.8 (GDP) & 0.48 \\
        Trp6 & 0.066 & 10.0 & 10.1 (GDP) & 0.83 \\
        Trp7 & 0.157 & 10.0 & 19.7 (GTP) & 1.52 \\
        Trp8 & 1.003 & 31.9 &  9.9 (GDP) & 3.87 \\
        \bottomrule
    \end{tabular}
\end{table}

Table~\ref{tab:structure} contextualises the site heterogeneity, and Fig.~\ref{fig:sasa_tau_mobility} relates these structural descriptors to the fitted bath parameters. The amplitude structure is more regular than the timescales. The water-reorientation amplitude $\sigma_2$ tracks SASA closely (Pearson $r = 0.93$): more exposed indoles sense a larger fluctuating aqueous field, as expected. The libration amplitude $\sigma_1$ follows the same trend more weakly ($r = 0.82$). The protein-conformational amplitude $\sigma_3$ is instead non-monotonic (V-shaped): both deeply buried residues (Trp5, Trp3, Trp6) and the most exposed one (Trp8) carry large slow-mode variance, while semi-exposed sites (Trp7, Trp4) are minimal. Buried residues directly sense slow packing fluctuations of the surrounding protein; exposed residues sense slow loop and surface rearrangements; the intermediate-SASA sites sit in locally rigid pockets that suppress the nanosecond mode. Trp8 is the outlier on every axis: highest $\sigma$, largest SASA, highest mobility, closest to a nucleotide, most hydrated.

The relaxation timescales $T_k$ show no comparably clean structural trend ($|r| < 0.44$ for all $T_k$--SASA pairs) and span wide ranges, especially $T_3$ (0.45--8.5~ns, a factor of ${\sim}20$). The site-by-site variation and its physical interpretation, including the poor separation of $T_2$ and $T_3$ at Trp6 and the poorly constrained $T_3$ at Trp4, are discussed in Sec.~\ref{sec:appendix-tau-heterogeneity}.

\paragraph{Why $\sigma_{\mathrm{short}}$ can exceed $\sigma$.}
For most sites the long-trajectory standard deviation exceeds the short-trajectory one ($\sigma > \sigma_{\mathrm{short}}$, Table~\ref{tab:bath_summary}), because the 2~ns short trajectory undersamples the nanosecond slow mode. The sign of the difference is governed by whether $T_3$ is resolved within the 2~ns window. The two most buried sites are exceptions: Trp5 (SASA~$= 0.029$~nm\textsuperscript{2}, $T_3 = 0.96$~ns) and Trp6 (SASA~$= 0.066$~nm\textsuperscript{2}, $T_3 = 0.49$~ns) have slow modes that relax well inside the 2~ns short trajectory ($2~\mathrm{ns}/T_3 \approx 2$ and $4$, respectively), so there is no missing slow-mode variance to bias $\sigma_{\mathrm{short}}$ downward. The two estimates agree for Trp5 ($\sigma_{\mathrm{short}}/\sigma = 1.01$) and the short slightly exceeds the long for Trp6 ($\sigma_{\mathrm{short}}/\sigma = 1.12$), reflecting sampling variability of a slow mode sampled only a few times within the window. Trp8 ($T_3 = 1.36$~ns, $\sigma_{\mathrm{short}}/\sigma = 1.04$) follows the same logic. In contrast, sites with $T_3 \gg 2$~ns, such as Trp3 ($T_3 = 6.8$~ns, $\sigma_{\mathrm{short}}/\sigma = 0.79$) and Trp4 ($T_3 = 8.5$~ns, $\sigma_{\mathrm{short}}/\sigma = 0.86$), show the expected $\sigma > \sigma_{\mathrm{short}}$ because the short trajectory captures only a fraction of a slow-mode period.

\section{Site-to-site variation of relaxation timescales}
\label{sec:appendix-tau-heterogeneity}

\begin{figure}[t]
    \centering
    \includegraphics[width=\columnwidth]{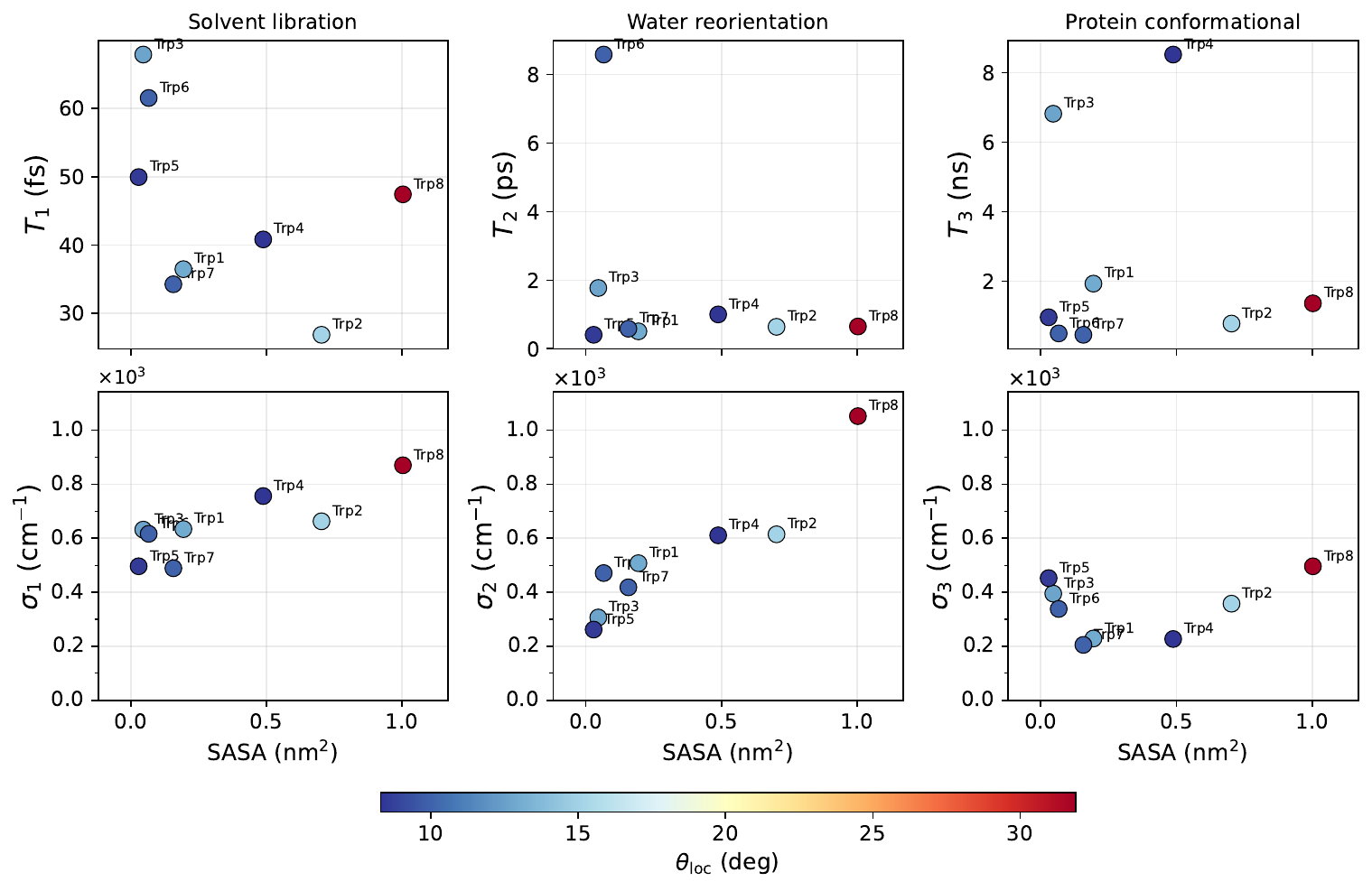}
    \caption{Bath parameters vs.\ whole-residue SASA for each Trp site, coloured by the local dipole angular deviation $\theta_{\mathrm{loc}}$ (Table~\ref{tab:structure}). \textbf{(top)} Relaxation timescales $T_k$: $T_1$ (solvent libration, 27--68~fs), $T_2$ (water reorientation, 0.4--8.6~ps), $T_3$ (protein conformational, 0.5--8.5~ns), showing no clean structural trend ($|r|<0.44$). \textbf{(bottom)} Per-component amplitudes $\sigma_k = \sigma_m\sqrt{f_k}$: $\sigma_2$ correlates strongly with SASA ($r=0.93$), $\sigma_1$ more weakly ($r=0.82$), while $\sigma_3$ is non-monotonic (V-shaped), with both deeply buried and highly exposed sites carrying large slow-mode variance.}
    \label{fig:sasa_tau_mobility}
\end{figure}

Figure~\ref{fig:sasa_tau_mobility} visualises the wide site-to-site variation in the three fitted timescales. The colour encodes $\theta_{\mathrm{loc}}$, the local angular mobility of the Trp difference dipole. For each MD frame, the global protein rotation $R(t)$ is removed by solving Wahba's problem (optimal rotation mapping the 8-vector dipole set to its time-averaged reference, via SVD). The corrected dipole directions $\hat{\bm{n}}_m^{\mathrm{loc}}(t)$ are then compared to their own time average, giving the mean angular tilt $\theta_m = \langle\arccos(\hat{\bm{n}}_m^{\mathrm{loc}}(t)\cdot\langle\hat{\bm{n}}_m^{\mathrm{loc}}\rangle)\rangle$. This is the orientational analogue of the root-mean-square fluctuation: it measures how many degrees the dipole typically deviates from its mean orientation, independent of whole-protein tumbling. Low $\theta_{\mathrm{loc}}$ (blue) indicates a rigidly held dipole; high $\theta_{\mathrm{loc}}$ (red) indicates conformational flexibility.

The variation in $T_k$ is real and reflects the distinct local environment of each Trp, but individual values carry varying degrees of uncertainty.

\paragraph{$T_3$ (protein conformational).}
The largest variation spans an order of magnitude, from 0.45~ns (Trp7) to 8.52~ns (Trp4). Deeply buried residues in rigid regions can sense very slow conformational modes: Trp3 (SASA~$= 0.046$~nm\textsuperscript{2}, $T_3 = 6.8$~ns, $f_3 = 0.24$) is the clearest example. However, burial alone does not determine $T_3$: Trp5, the most buried site (SASA~$= 0.029$~nm\textsuperscript{2}), has $T_3 = 0.96$~ns. Sites near the $\alpha$--$\beta$ interface (Trp4, Trp7) show contrasting behaviour: Trp4 has $T_3 = 8.52$~ns but only $f_3 = 0.05$, meaning the slow component contributes just 5\% of the total variance. Because the 40~ns trajectory spans only $T_{\mathrm{traj}}/T_3 \approx 4.7$ slow-mode periods, this value is poorly constrained; the true $T_3$ could lie anywhere in the 5--15~ns range. Trp3 ($T_3 = 6.8$~ns, $T_{\mathrm{traj}}/T_3 \approx 5.9$) is similarly marginal. All other sites have $T_{\mathrm{traj}}/T_3 > 20$ and are well constrained.

\paragraph{$T_2$ (water reorientation).}
Most sites cluster between 0.4 and 1.0~ps, consistent with hydrogen-bond reformation dynamics. Trp6 ($T_2 = 8.58$~ps) is a pronounced outlier. Its $T_3/T_2$ ratio is only 57, compared to $>700$ for every other site, indicating that the $T_2$ and $T_3$ components are poorly separated in the tri-exponential fit. Trp6 is buried (SASA~$= 0.066$~nm\textsuperscript{2}) and close to GDP (10.1~\AA); the anomalous $T_2$ likely reflects an intermediate timescale from nucleotide-associated ordered water or phosphate-group fluctuations that the fit cannot cleanly assign to either the $T_2$ or $T_3$ component.

\paragraph{$T_1$ (solvent libration).}
The range is modest (27--68~fs, a factor of ${\sim}2.5$). The short trajectory's 10~fs cadence resolves $T_1$ at 3--7 frames, so absolute values carry ${\sim}30\%$ uncertainty for the fastest sites (Trp1, 2, 7). Nevertheless, the ordering is physically meaningful: buried sites (Trp3, Trp6) tend to have slower $T_1$, possibly reflecting non-water contributions to the fast mode from backbone fluctuations.

\paragraph{Implications for the exciton dynamics.}
Despite these uncertainties, the conclusions of Sec.~\ref{sec:results-exciton} are robust because they depend only on the \emph{order of magnitude} of $T_k$, not on precise values. All sites satisfy $\sigma/|J| \gg 1$ (strong disorder), all Kubo numbers $\kappa_k \gg 1$ (non-Markovian), and even the shortest $T_3$ (0.45~ns) exceeds $\hbar/J \approx 90$~fs by three orders of magnitude. The site-to-site heterogeneity is itself a physical feature: no single bath parameter set can describe all eight Trps, reinforcing the need for per-site noise modelling.

\section{Eigenstate localisation of the 8-site Hamiltonian}
\label{sec:appendix-pr}

\begin{figure}[H]
    \centering
    \includegraphics[width=\columnwidth]{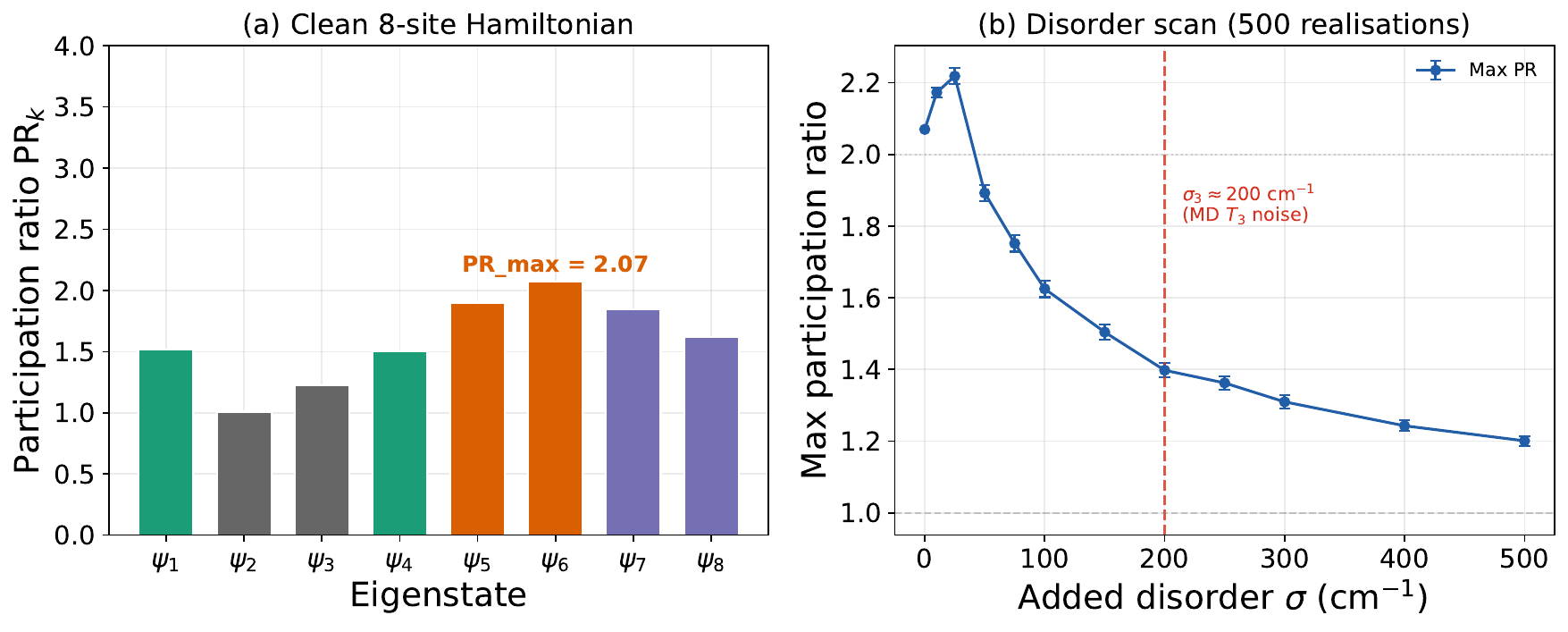}
    \caption{Eigenstate localisation in the 8-site Trp Hamiltonian, quantified by the participation ratio $\mathrm{PR}_k = 1/\sum_i|c_i^{(k)}|^4$. \textbf{(a)} PR of each eigenstate of the clean Hamiltonian. No eigenstate exceeds PR~$= 2.07$ (Trp4--Trp7 pair), because the 388~cm\textsuperscript{-1} spread of site energies overwhelms the couplings ($|J|_{\max} = 59$~cm\textsuperscript{-1}). \textbf{(b)} Maximum PR across all eigenstates vs.\ added Gaussian disorder $\sigma$ (500 realisations per point). The slight initial rise at small $\sigma$ ($\lesssim 25$~cm\textsuperscript{-1}) is a disorder-assisted resonance effect: small random shifts can occasionally reduce the intrinsic detuning between coupled pairs (e.g.\ $\Delta\epsilon_{47} = 41$~cm\textsuperscript{-1}). At larger $\sigma$, disorder dominates and localisation sets in. The vertical dashed line marks the MD-derived $T_3$ noise level ($\sigma_3 \approx 200$~cm\textsuperscript{-1}); at this point the max PR drops to 1.40.}
    \label{fig:pr_scan}
\end{figure}

Diagonalising the clean Hamiltonian (Table~\ref{tab:hamiltonian}) reveals that the eight-site network fragments into three weakly connected pairs (Trp4--Trp7, Trp6--Trp8, Trp2--Trp3) and two nearly isolated sites (Trp1, Trp5). The participation ratio $\mathrm{PR}_k = 1/\sum_i|c_i^{(k)}|^4$ measures the effective number of sites spanned by eigenstate $k$: it equals 1 for a state fully localised on a single site and $N$ for a state uniformly delocalised across all $N$ sites. As shown in Fig.~\ref{fig:pr_scan}(a), no eigenstate exceeds PR~$= 2.07$, far below the fully delocalised limit of 8. The root cause is the large spread of diagonal site energies (388~cm\textsuperscript{-1}), which exceeds the strongest coupling by a factor of 6.6.

Adding Gaussian static disorder $\sigma$ to the diagonal further localises the eigenstates [Fig.~\ref{fig:pr_scan}(b)]. At the MD-derived $T_3$ level ($\sigma_3 \approx 200$~cm\textsuperscript{-1}), the maximum PR drops to 1.40, indicating near-complete single-site confinement under the realistic bath. These results are referenced in the superradiance discussion of Sec.~\ref{sec:discussion}.

\end{document}